\documentclass[10pt,conference]{IEEEtran}

\usepackage[utf8]{inputenc} 
\usepackage[T1]{fontenc}
\usepackage{lineno}
\usepackage{graphicx}%
\usepackage{multirow}%
\usepackage{amsmath,amssymb,amsfonts}%
\usepackage{textcomp}%
\usepackage{manyfoot}%
\usepackage{booktabs}%
\usepackage{algorithm}%
\usepackage{algorithmicx}%
\usepackage{algpseudocode}%
\usepackage{listings}%
\usepackage{fix-cm}
\usepackage{natbib}
\usepackage{url}
\usepackage{tabularx}
\usepackage{xurl}
\usepackage[table]{xcolor}%

\usepackage[inkscapearea=page]{svg}

\usepackage[most]{tcolorbox}   
\usepackage{lipsum}            

\newtcolorbox{myanswerbox}[1][]{
    enhanced,                     
    colback=gray!5,                
    colframe=black,              
    boxrule=0.8pt,                
    arc=4pt,                       
    outer arc=4pt,
    left=6pt, right=6pt, top=6pt, bottom=6pt,
    fonttitle=\bfseries,          
    #1                            
}

\begin{document}

\title{Crossing Margins: Intersectional Users' Ethical Concerns about Software}

\author{Lauren Olson \and
        Tom P. Humbert \and
        Ricarda Anna-Lena Fischer\and
        Bob Westerveld \and
        Florian Kunneman \and
        Emitzá Guzmán}



\maketitle

\begin{abstract}
Many modern software applications present numerous \textit{ethical concerns} due to conflicts between users' values and companies' priorities. Intersectional communities, those with multiple marginalized identities, are disproportionately affected by these ethical issues, leading to legal, financial, and reputational consequences for software companies, as well as real-world harm for intersectional users. Historically, the voices of intersectional communities have been systematically marginalized and excluded from contributing their unique perspectives to software design, perpetuating software-related ethical concerns.

This work aims to fill the gap in research on intersectional users' software-related perspectives and provide software practitioners with a methodology for analyzing intersectional voices in software ethics discourse. We collected 36,777 posts from over 700 intersectional subreddits discussing software applications and utilized large language models to identify ethical concerns in these posts. We then applied regression models with counterfactual analysis to examine how intersectional identity dimensions shape the amplification or suppression of ethical concern expression across software genres, and conducted a time-series analysis to examine how concern expression varies over time in relation to real-world events. As a case study in the social media domain, we further demonstrate how identified ethical concerns can be prioritized to surface issues warranting timely developer attention, validated against survey-derived ground truth. Together, these analyses form the basis of a nascent feedback-driven framework for assessing whether software systems are meeting the needs of intersectional users. 
  
\end{abstract}


\section{Introduction}
Software applications play a substantial role in shaping how individuals communicate, work, and engage with the world, influencing nearly every aspect of daily life. Because software is continually updated in response to user needs and reported issues, feedback plays a critical role in shaping the direction of its design, functionality, and inclusivity. Prior research indicates that \textbf{user feedback} for software applications is predominantly provided by middle-aged men~\citep{tizard2022voice}, and that feedback exhibits significant variation based on demographic factors~\citep{fischer2021does}. This demographic skew means that the values, needs, and harms experienced by a broader, more diverse user base, particularly multiply marginalized or \textit{intersectional} users, are often excluded from the design and evolution of software systems.

\textit{Intersectionality}\footnote{Although some work uses ``intersectionality'' to mean a generic two-variable analysis, we adopt Kimberl\'e Crenshaw's original formulation~\citep{crenshaw2013demarginalizing}.}, a concept coined by Kimberl\'e Crenshaw in 1989, describes how overlapping forms of marginalization compound to create unique and intensified experiences of discrimination~\citep{crenshaw2013demarginalizing}. Intersectional communities, \textbf{those who belong to multiple \textit{marginalized} groups}, such as Black women, queer disabled users, or neurodivergent people from the Global South, face compounded forms of harm when their experiences are left out of feedback and development cycles~\citep{buolamwini2018gender, costanza2020design}. Feedback from these communities is vital but often missing in the software development lifecycle. When absent, ethical issues such as \textit{algorithmic discrimination, inaccessibility, and identity erasure} become more likely and more harmful. This exclusion can have not only social consequences but also legal and financial ones, as seen in cases where platforms were held accountable for discriminatory outcomes~\citep{Benner_Thrush_Isaac_2019, Biron_2022, Clayton_2021, Thorbecke_2020, Whitcomb_2017}.

Historically, the voices of intersectional communities have been marginalized~\citep{crenshaw2013demarginalizing, crenshaw2013mapping}, both in software industry practices and academic research. Studies have begun to surface ethical issues that disproportionately affect these communities: facial analysis systems have shown error rates 20--35\% higher for darker-skinned women than for lighter-skinned men~\citep{buolamwini2018gender}; generative language models reinforce occupational stereotypes about women of color and queer individuals~\citep{kirk2021bias}; and low-income Black Americans, racial minority women, and trans individuals face exclusion in domains such as health technology, maternal care apps, and online identity systems~\citep{kim2022designing, oguamanam2023intersectional, oakley2016disturbing}. While these studies offer essential insights into specific user populations, they are typically small in scale, limited in scope, and largely descriptive, identifying where harm occurs without examining how intersectional identity shapes the expression of those harms, how those harms shift over time, or how they might be systematically surfaced for developer attention.

These three gaps directly motivate our research. First, while we know that intersectional users face compounded harm, intersectionality theory predicts that overlapping marginalized identities produce amplified or suppressed experiences that are not reducible to the sum of their parts~\citep{crenshaw2013demarginalizing}; yet we do not yet understand whether this amplification and suppression extends to the \textit{type} of ethical concerns users raise about software: whether certain communities disproportionately express concerns around privacy, misinformation, or cyberbullying, and whether those patterns hold when controlling for other community-level factors. Second, ethical concern expression does not occur in a vacuum: world events, platform controversies, and social movements plausibly shape when and how concerns are raised, yet this temporal dimension remains largely unexamined. Third, even where concerns are visible in user-generated data, there is no established approach for systematically prioritizing them in ways that reflect what intersectional users themselves consider most urgent, a gap that is particularly acute in the social media domain, where harms are well-documented but developer accountability mechanisms remain weak.

This gap motivates a shift toward \textit{feedback-driven auditing}: approaches that use real user-generated signals to assess whether software systems are meeting, or failing, the needs of their users. Recent work has demonstrated that how organizations handle user feedback has measurable downstream consequences: omitting feedback during update cycles can substantially depress user ratings~\citep{gokgoz2025if}, and ethically relevant concerns that go unaddressed likely compound harm over time. This study takes a step toward such a pipeline by systematically analyzing software-related ethical concerns raised across over 700 intersectional subreddits on Reddit. We adopt the definition of \textit{ethical concern} from Olson et al.~\citep{Olson2023} as ``\textit{a worry or care the user or their group faces about what is right or good}'' within the software platform, and address the following research questions:

\vspace{1mm}
\noindent \textbf{(RQ1) Intersectional Expression:} How does intersectionality shape how ethical concerns about software are expressed on Reddit?\\
\noindent \textbf{(RQ2) Time:} How do intersectional users' ethical concern expressions vary over time on Reddit, and what real-world events drive those patterns?\\
\noindent \textbf{(RQ3) Prioritization} \textit{(Social Media Case Study):} In the social media domain, how can ethical concern prioritization surface concerns from Reddit warranting timely developer attention?
\vspace{2mm}

Together, these analyses form the basis of a nascent auditing framework: one that identifies what harms are present and how intersectional identity shapes their expression (RQ1), tracks how concerns evolve in response to world events (RQ2), and demonstrates how high-priority concerns can be surfaced for developer attention in a specific high-harm domain (RQ3). To support these analyses, we use a novel dataset of 36,777 Reddit posts from intersectional communities mentioning the top 50 software applications by user count, including 6906 posts explicitly discussing specific ethical concerns. We conduct an ethical concern category classification, intersectional regression analysis with counterfactual probing, time-series analysis, and concern prioritization using a combination of LLMs, sentiment analysis, and anomaly detection, complemented by a survey of 102 participants to validate prioritized concerns against user-reported ground truth. The specific contributions are:

\begin{itemize}
    \item A dataset of 36,777 Reddit posts from over 700 intersectional subreddits that mention the top 50 software applications by user count, of which 6906 posts explicitly address ethical concerns.

    \item An identification and categorization of ethical concern types across 33 intersectional communities, with regression models and counterfactual analysis to assess how community-level intersectional factors shape ethical concern expression (RQ1).

    \item A 7-year time series analysis linking ethical concern posts to broader temporal patterns and world events, offering contextual insight into when and how concerns are raised (RQ2).

    \item A prioritization approach combining LLMs, sentiment analysis, and anomaly detection to surface social media ethical concerns warranting timely developer attention, validated against survey-derived ground truth from 102 participants (RQ3).

    \item A proof-of-concept auditing framework integrating the above components to assess whether software systems are surfacing and responding to the ethical concerns of marginalized users.
\end{itemize}

All data, code, and annotation guidelines are made available in our replication package.\footnote{\url{www.doi.org/10.6084/m9.figshare.26888128}}

\vspace{2mm}
\noindent\textbf{Content Warning:} This paper discusses sensitive topics, including miscarriage, suicide, body image, transphobia, drug use, and misogyny.
\section{Related Work}
\textbf{Mining User Feedback.}
Users provide critical feedback for software development on many platforms, including app stores~\citep{Dennis2013, hoon2013analysis, guzman2018user, fischer2021does}, Twitter~\citep{guzman2016needle, williams2017mining, nayebi2018app, tabbassum2023towards}, and Reddit~\citep{iqbal2021mining, Olson2023}. A systematic review by Dabrowski et al.~\citep{dkabrowski2022analysing} identified nine primary techniques for user feedback analysis, including classification, prioritization (recommendation), and content analysis.

Notably, missing from this list is time series analysis due to its nascent stage in the field~\citep{dkabrowski2022analysing}. In fact, only a few studies have used time series {analyzes} to analyze feedback over time~\citep{gao2018online, stronstads}; specifically, to detect thematic trends~\citep{gao2018online} and anomalies~\citep{stronstads, gao2018online}. In our work, we use time series analysis to examine the evolution of intersectional communities' discussions about ethical concerns in software.

Previous work has used traditional machine learning algorithms~\citep{guzman2015ensemble, guzman2017little, williams2017mining, maalej2015bug, panichella2015can} as well as deep learning techniques~\citep{li2022narratives, shahin2023study, Tjikhoeri2024} to classify user feedback. Prioritisation systems for app‑store reviews typically combine internal variables (e.g., comment volume, rating, sentiment, recency) \citep{9952173,chen2014ar,licorish2017attributes,malgaonkar2022prioritizing} with external variables (e.g., popularity, social rank) when dealing with social‑media feedback \citep{guzman2017little}. Evaluation is often performed with small external samples (often fewer than 20 practitioners) and without demographic reporting \citep{guzman2017little}. In contrast, our evaluation gathers feedback from more than 100 participants, with at least 50\% belonging to intersectional groups, and we supplement internal metrics with popularity, entropy, sentiment, and recency.

\textbf{User Feedback related to Ethical Concerns.}
Research has explored the identification of ethical concerns \citep{tushev2020digital, Tjikhoeri2024, khalid2014mobile, besmer2020investigating} and violations of human values \citep{obie2021first, shams2020society} within user feedback on software systems. Reddit has emerged as a critical platform for this analysis due to its support for extended posts, anonymity, and community-focused interactions \citep{Olson2023}. Prior studies validate Reddit's utility as a source of user feedback, finding that approximately 54\% of posts in \textit{software-related} subreddits provide actionable insights for software evolution \citep{iqbal2021mining}.

Existing research has leveraged Reddit to probe into user-expressed ethical concerns~\citep{li2022narratives, Olson2023}. For instance, one study highlighted that discussions often revolve around privacy issues, including software policies and permissions~\citep{li2022narratives}. Another investigated ethical concerns among marginalized communities, finding discrimination and misrepresentation to be particularly relevant concerns~\citep{Olson2023}. However, this study examines user preferences across single marginalized identities, while we explore feedback from users with \textit{multiple} marginalized identities. By focusing on a single identity or none at all, the generalizability of findings is critically restricted. Our study aims to bridge this gap by providing a longitudinal analysis of intersectional ethical concerns, linking them to global events, and formulating a user-centric prioritization framework that captures the varied and complex views of intersectional communities \citep{Olson2023}.

\textbf{Intersectional Communities.}
Prior research on integrating intersectional user perspectives into software products has employed methods such as constructing intersectional personas \citep{mendez2019gendermag}, devising theoretical frameworks \citep{rankin2019straighten, erete2018intersectional, foulds2020intersectional, mcdonald2020privacy, sanchez2021framework, klumbyte2022critical}, and conducting focused user studies to meet the specific needs of intersectional communities \citep{kim2022designing, hedditch2023design, kumar2020taking, wong2018designing, oguamanam2023intersectional, moitra2021negotiating, andalibi2022lgbtq}. Despite their utility, the use of intersectional personas has been criticized for reinforcing stereotypical identity views \citep{marsden2016stereotypes}. Alternately, we do not use abstractions; we capture greater internal diversity and reduce the risk of misrepresentation by using real data.

Frameworks designed to facilitate user-centered research have yielded insights into the preferences of diverse intersectional populations, such as Black perinatal women, Black youth, Black COVID patients, Indian women, Indian queer individuals, and gender non-conforming users, across various software applications \citep{kim2022designing, hedditch2023design, kumar2020taking, wong2018designing, oguamanam2023intersectional, moitra2021negotiating, andalibi2022lgbtq}. For instance, research involving LGBTQ+ individuals experiencing recent pregnancy losses revealed significant challenges in finding online support due to intersecting marginalized identities \citep{andalibi2022lgbtq}. Our work seeks to expand on these findings, by providing a more general overview of intersectional users needs and concerns when it comes to software.

Additionally, another line of inquiry has examined intersectional biases in software systems, particularly those leveraging machine learning \citep{buolamwini2018gender, cabrera2019fairvis, steed2021image} and natural language processing technologies \citep{kirk2021bias, guo2021detecting, ghai2021wordbias, tan2019assessing}. These studies have often found pronounced biases against intersectional groups, often more severe than those affecting singly marginalized identities \citep{buolamwini2018gender}. Highlighting these disparities is crucial, and our research contributes to this effort by providing an in-depth analysis of the diverse and complex experiences of intersectional groups, aiming to challenge and reduce existing stereotypes in software engineering.

\section{Scope and Data}
\begin{figure}
    \centering
    \includegraphics[width=.5\textwidth]{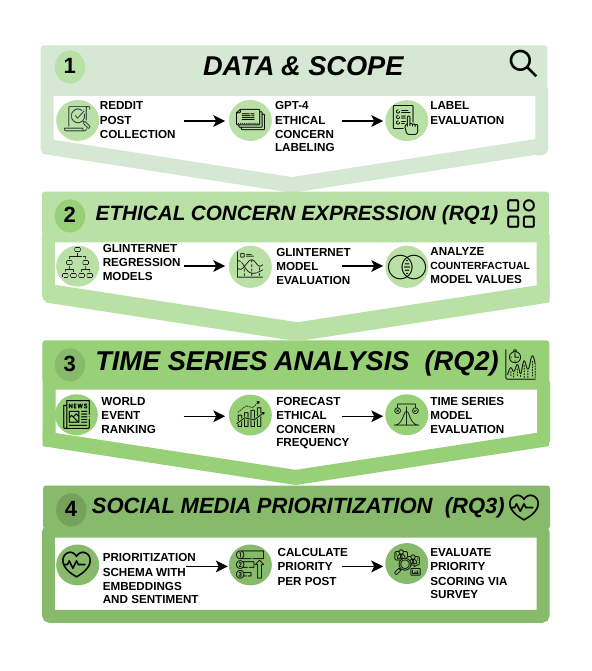}
\caption{Overview of the study methodology, comprising four stages: (1) data collection via community detection across 745 subreddits and scraping of 36,777 app-related posts; (2) ethical concern (EC) categorization using GPT-4 multi-class labeling with label evaluation and intersectional regression analysis (RQ1); (3) time series analysis forecasting EC frequency across the full post corpus with model evaluation (RQ2); and (4) a social media prioritization case study computing priority scores for 4,231 EC posts and validating against ground truth labels from 102 survey participants (RQ3). }
    \label{fig:methods}
\end{figure}

The scope of this study is categorized into two distinct analytical phases based on temporal relevance and research objectives. While the initial phase led to a dataset of 36,777 thousand posts that mention a software application, the second phase focuses on ethical concerns within the social media domain (see Figure~\ref{fig:methods} for an overview).

\subsection{Intersectional Communities} 
We chose seven marginalized communities to detect intersections between, namely Black, indigenous, and people of color (BIPOC), women or people assigned female at birth (AFAB), LGBTQIA+, people of lower socio-economic status (SES), people from the Global South, people that have physical health issues and people that have mental health differences (includes neurodivergent communities)~\citep{costanza2020design}. We used a comprehensive list of 586 marginalized subreddits~\citep{Olson2023} as the seed list for identifying intersectional subreddits. A subreddit was considered intersectional if its description indicated that members belonged to at least two of the seven marginalized communities. We searched each subreddit from our seed list on Reddit, using the `communities' results tab to manually identify additional relevant subreddits. Intersectional communities were often evident from subreddit names, such as r/BlackWomenDivest,' which intersects BIPOC' and women/AFAB.' In some cases, closer evaluation was needed, such as with r/ketogd,' where members manage gestational diabetes, a condition affecting pregnant individuals. This subreddit was categorized under women/AFAB' and `physical health.' Repeating this process for each subreddit, we compiled 746 intersectional subreddits, included in the replication package.

\begin{table}
\footnotesize
\caption{Original Intersectional Community Groups and number of subreddits associated to each}
\rowcolors{2}{gray!15}{white}
\begin{tabular}{lp{6.5cm}l}
\hline
& \textbf{Intersectional Communities}  &\textbf{\#}\\\hline
1&  BIPOC $\times$ Lower SES&1\\
2&  Physical Health $\times$ Global South&1\\
3&  Lower SES $\times$ Global South&1\\
4&  Women/AFAB $\times$ Mental Health $\times$ Global South&1\\
5&  Women/AFAB $\times$ Lower SES&2\\
6&  Lower SES $\times$ LGBTQIA+&2\\
7&  BIPOC $\times$ Women/AFAB $\times$ Mental Health&2\\
8&  Physical Health $\times$ Lower SES&2\\
9&  BIPOC $\times$ Physical Health $\times$ Global South&2\\
10&  Women/AFAB $\times$ Lower SES $\times$ LGBTQIA+&3\\
11&  Women/AFAB $\times$ Global South $\times$ LGBTQIA+&3\\
12&  Physical Health $\times$ LGBTQIA+&4\\
13&  BIPOC $\times$ Women/AFAB $\times$ Physical Health $\times$ Global South&4\\
14&  BIPOC $\times$ Mental Health&5\\
15&  Mental Health $\times$ Physical Health&6\\
16&  BIPOC $\times$ Global South&10\\
17&  Women/AFAB $\times$ Physical Health $\times$ LGBTQIA+&11\\
18&  Global South $\times$ LGBTQIA+&12\\
19&  Women/AFAB $\times$ Mental Health $\times$ Physical Health&13\\
20&  BIPOC $\times$ Women/AFAB $\times$ Global South $\times$ LGBTQIA+&13\\
21&  Women/AFAB $\times$ Global South&13\\
22&  BIPOC $\times$ LGBTQIA+&14\\
23&  Mental Health $\times$ Global South&15\\
24&  BIPOC $\times$ Mental Health $\times$ Global South&15\\
25&  Women/AFAB $\times$ Mental Health $\times$ LGBTQIA+&16\\
26&  BIPOC $\times$ Women/AFAB $\times$ LGBTQIA+&18\\
27&  BIPOC $\times$ Women/AFAB $\times$ Global South&20\\
28&  Mental Health $\times$ LGBTQIA+&21\\
29&  Women/AFAB $\times$ Mental Health&29\\
30&  BIPOC $\times$ Women/AFAB&31\\
31&  BIPOC $\times$ Global South $\times$ LGBTQIA+&35\\
32&  Women/AFAB $\times$ Physical Health&82\\
33&  Women/AFAB $\times$ LGBTQIA+&338\\ \hline
\end{tabular}
\label{table:og_intersectional}
\end{table}

 We identified 33 unique intersectional communities in our dataset (see Table~\ref{table:og_intersectional}).

\subsection{Software Applications} The scope includes the top 50 US mobile applications by user count (data.ai).\footnote{included in the replication package}   We chose the US as the geographic domain, as Reddit's primary user base is US-centered~\citep{reddit_2022}. We added the word ``app" after app names that are homonymous with often-used words to avoid false positives (e.g., ``McDonalds $\rightarrow$ McDonalds app").

\subsubsection{Time} Due to a limitation of Reddit's API, which restricts users to retrieving only the most recent 1000 posts per subreddit, our scraper collected up to 1000 posts (or fewer if the subreddit contained less than 1000) mentioning our app names from each intersectional subreddit. In total, we collected 36,777 posts on June 20, 2023. These posts span 14 years, from 2009 to 2023, with 2021 serving as the median date.

\subsection{Ethics and Privacy}
We collected only publicly available Reddit data, adhering to Reddit’s terms of service\footnote{\url{https://www.redditinc.com/policies/user-agreement-april-18-2023} }, using a custom PRAW~\citep{Boe_2016} scraper. To safeguard user privacy, all posts in the paper are paraphrased using the Google test\footnote{googling the text to see if the search engine identified the original user post} to prevent traceability, and identifying details, including usernames, timestamps, and subreddit names, were removed. These measures were independently verified by a second author. We did not obtain individual user consent due to practical constraints and concerns about influencing natural behavior. While we considered seeking moderator consent, we determined it was not a suitable proxy for user consent, given the informal and unelected nature of subreddit governance.

\subsection{Ethical Concern Labeling}
To identify the ethical concerns raised by intersectional communities, we conducted this classification process in two phases. In the first phase, we filtered the dataset to include only posts after 2017 from the top six most-mentioned social media platforms. The 2017 cutoff was chosen because it represented a five-year window at the time of the study, ensuring that the issues discussed were relatively recent and that annotators were likely familiar with them from prior annotation work on ethical concerns in software contexts. Social media platforms were prioritized for this phase because our annotation team had accumulated substantial familiarity with this domain through repeated engagement with ethical concern labeling tasks, making them well-positioned to evaluate model accuracy in this subset. Using OpenAI’s GPT-4 model, we assigned each post one label from a pre-defined set of ethical concern categories derived from prior work (see Table \ref{table:taxonomy4})~\citep{Tjikhoeri2024}. At the time of access (August 2023), GPT-4 was OpenAI's highest-performing model on a range of tasks. We then manually reviewed ten posts per cateory to assess the accuracy of the model-generated labels and retained categories that achieved high validation scores. 

\subsubsection{Multi-class Labeling for Social Media Posts}
\label{sec:multiclass}
To classify the ethical concern type of the posts, we used the GPT-4 engine and an existing ethical concerns taxonomy~\citep{Tjikhoeri2024}.

\begin{table*}
\caption{Ethical Concerns Taxonomy (All examples from Tjikhoeri et al.~\citep{Tjikhoeri2024} with the exception of social isolation, which is from Olson et al.~\citep{Olson2023}))}
\centering
\footnotesize
\rowcolors{2}{gray!15}{white}

\begin{tabular}{lp{6.5 cm}p{6.5 cm}}
\toprule
\textbf{Ethical Concern}  & \textbf{Definition}     & \textbf{Example}   \\ \midrule 

\textit{Addiction}             & The application's design or content is addictive to the user. & I really hate the YouTube rabbit hole and \textbf{the design of YouTube as addictive}. It is a very big problem.  \\ 

\textit{Censorship}            & The application hides certain information, or the user's content or profiles are removed or demoted.   & This application propagates radical ideology and \textbf{suppresses open-mindedness and conservative views}. I cannot express how much censorship is illegal and is what I observe on this platform. \\ 

\textit{Cyberbullying}         & The application’s users exhibit intentional harmful or mean behavior towards the user. & There's still alot of hate etc on there. \textbf{Its sad that you don't care about bullies}  \\ 

\textit{Discrimination}        & The application or its community is participating in prejudicial treatment of different categories of people. & Racist and discriminating. They have been \textbf{proven to suppress black lives matter, disabled, poor and anyone that's not conventionally attractive.}  \\ 

\textit{Harmful Advertising}   & The application hosts advertisements that mislead or harm the user.            & Too many \textbf{advertising for gambling directed at kids}   \\ 

\textit{Inappropriate Content} & The application hosts content, including but not limited to posts, comments, or multimedia (not advertisements) that distresses the user. & Theres \textbf{so many sexual videos} they might as well call this pornhub Junior   \\ 

\textit{Misinformation}        & The application spreads false or inaccurate information to users. &   Amazing app it is my teacher but it just only one minor problem that \textbf{fake news is more than real news} \\ 

\textit{Privacy}               & The application and community does not keep the user’s information secure or uses it for non-consensual purposes. The application does not provide the user the ability to control access to their information. & The \textbf{amount of privacy you need to give up} to have this overhyped egg timer in your house is ridiculous. (Amazon Alexa) \\ 

\textit{Safety}                & The application or its community has caused physical or mental health issues or other safety risks for the user. &  Encourages young children do do \textbf{very dangerous tiktok `trends' like spraying your bathroom mirror with a flammable substance and setting it on fire.}    \\ 

\textit{Scam}                  & The application or its community has engaged in deceitful behavior to gain something, usually money or goods, from the user.  & Get \textbf{ scammers wanting money. And I have fell for a sob story.} Then the scammers delete their accounts  \\ 

\textit{Social Isolation}      & The application or its community causes the user to feel lonely. & That realization pushed me to \textbf{connect with people online, but it backfired.} I shared my interests, hoping to engage, but was met with negativity and \textbf{felt dismissed.} \\ \bottomrule
\end{tabular}
\label{table:taxonomy4}
\end{table*}

\begin{figure*}
\centering
\includegraphics[width=.7\textwidth]{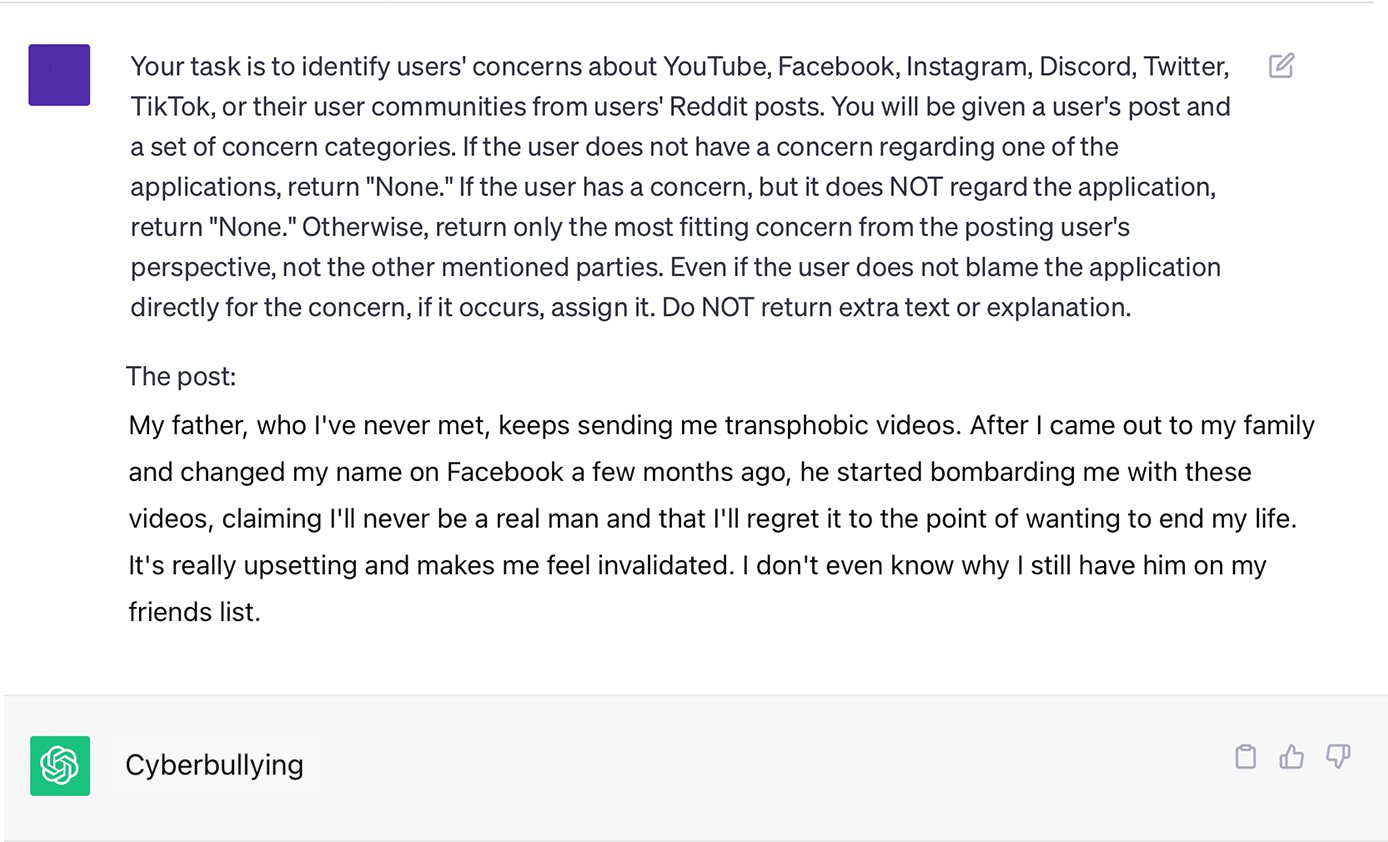}
\caption{Sample Prompt and Post to the GPT-4 engine to classify ethical concern types. The post is altered to maintain the author's privacy. For the study we used the API rather than the UI that is shown here for visualization purposes. Categories and their definitions were included at the end of the prompt (here excluded for the sake of readability).}
\label{fig:sample_prompt}
\end{figure*}

Each API request to GPT-4 (date accessed: August 2023) included a task prompt, definitions of the ethical concerns' categories, and an ethical concern post, windowed to focus the text (see Figure~\ref{fig:sample_prompt} for an example). We developed the task prompt by referencing and summarizing our original annotation guide's key points, which included the categories and definitions from previous work~\citep{Tjikhoeri2024, Olson2023}. The initial `set of user-informed ethical concerns' was developed by analyzing user feedback from the Google Play Store~\citep{Tjikhoeri2024}; it was later updated to reflect the ethical concerns present in Reddit data~\citep{Olson2023}. We used this updated version, tailored to Reddit data for our work (Table~\ref{table:taxonomy4}). Tjikhoeri et al.~\citep{Tjikhoeri2024}, through their coding approach, identified a single dominant ethical concern category. To streamline our analysis and focus on the primary ethical issues, we instruct the LLM to `return only the most relevant concern' (see Figure~\ref{fig:sample_prompt}).

We reduced this initial list after manually inspecting a sample of 400 posts; we removed \textit{content theft, identify theft, accessibility,} and \textit{accountability} as they did not occur in the sample. See Table~\ref{table:taxonomy4} for all considered categories. We also gave GPT-4 the option to label a post as `None.' In addition, we instructed GPT-4 to choose \textit{only one} label. 

\subsubsection{Label Evaluation for Social Media Posts}
We manually evaluated the quality of LLM-based categorization of ethical concerns. The first author validated ten classified posts from each category. If the post's GPT-4 ethical concern label fit the provided ethical concern definition, the author labeled the post as valid. For later analysis, we kept all categories that received an evaluation score of 8/10 or higher (social isolation=10/10, scam=9/10, privacy=9/10, misinformation=9/10, inappropriate content=9/10, harmful advertising=10/10, cyberbullying=9/10, addiction=6/10). We found that GPT-4 was too sensitive when labeling for addiction and tended to overlabel posts in this category. For example, GPT-4 labeled this post as addiction: `I want my partner's attention 24-7 and get super sad if he's scrolling on TikTok next to me,' confusing the poster's constant need for attention with scrolling.

\subsubsection{Full Dataset Classification via Fine-Tuned Model}

The initial EC categorization was conducted on a subset of social media posts, which limited the scope of our analysis. To extend classification to all 36,777 posts, we required a cost-effective solution, as re-labeling with GPT-4 at the same scale in March 2026 was not feasible due to budget constraints.

 GPT-4o mini achieved approximately 50\% accuracy at under \$5, and an initial fine-tuning attempt on GPT-4.1 nano (via with prompt adjustments and learning rate and epoch tuning) yielded approximately 56\% accuracy at a similar cost. We therefore fine-tuned GPT-4.1 mini on the existing labeled data (originally classified in August 2023), which produced a validation accuracy of 76.12\% at a cost of approximately \$20--\$30 USD.

\subsection{Ethical Concern Categorization Setup (RQ1)}
\label{sec:multilabels}

\subsubsection{Statistical Modeling}
To examine how intersectional social identities relate to ethical concerns about software, we followed a methodological framework tailored for analysing intersectional data~\citep{mccabe2025estimating, nielsen2025intersectional}: (1) we employed the group‑lasso interaction network (glinternet~\footnote{\url{https://www.rdocumentation.org/packages/glinternet/versions/1.0.12/topics/glinternet}}) \citep{lim2015learning} for binary outcomes (e.g., censorship present/absent); (2) we derived prevalence ratios (PRs) and prevalence‑ratio interactions (PRIs) from counterfactual predictions; and (3) we defined minimum important differences (MIDs) for PRs/PRIs $(\geq 1.22 or \leq 0.82)$ in line with previous work \citep{olivier2017relative}.

This approach is tailored to intersectional data as group-lasso interaction network extends traditional LASSO regression~\citep{tibshirani1996regression} to handle sparsity in high-dimensional data while preserving hierarchical interactions. The regularization penalty automatically shrinks irrelevant main effects and interactions toward zero, reducing overfitting when many demographic factors exist but some ethical concerns are rare. Hierarchical grouping ensures that interaction effects (e.g., BIPOC $\times$ LGBTQIA+) are considered only if their constituent main effects (e.g., BIPOC \textit{and} LGBTQIA+) are retained, reflecting the principle that an intersectional effect cannot exist without effects for the individual identities~\citep{mccabe2025estimating,nielsen2025intersectional}.

From the fitted models, we generated predicted prevalences from the model coefficients and conducted counterfactual analyses to assess whether predictions persist when specific demographic characteristics are hypothetically removed. These counterfactuals form the basis of prevalence ratios (PRs) for individual identities and prevalence ratio interactions (PRIs) for identity pairs. PRIs quantify whether the joint prevalence of two intersecting identities exceeds the additive expectation of each factor alone, allowing detection of amplifying or suppressing intersectional effects, another key principle of intersectional theory~\citep{crenshaw2013demarginalizing, crenshaw2013mapping}.

The only methodological extension we introduce is the use of the area under the ROC curve (AUC) to ensure model validity. After fitting the hierarchical regression models, we retained only those whose discrimination exceeded $AUC > 0.7$  (see Table \ref{tab:auc}), as models with $AUC > 0.7$ are considered to have an acceptable level of discrimination~\citep{hosmer2013applied}.

\begin{table}
\caption{Discriminative performance (AUC) of the hierarchical regression models for each ethical concern.}
\footnotesize
\centering
\rowcolors{2}{gray!15}{white}
\begin{tabular}{cc}
\toprule
\textbf{Concern}               & \textbf{AUC}   \\\midrule
Censorship            & 0.685 \\
Cyberbullying         & 0.640 \\
Discrimination        & 0.700 \\
Harmful Advertising   & 0.760 \\
Inappropriate Content & 0.678 \\
Misinformation        & 0.779 \\
Privacy               & 0.714 \\
Safety                & 0.695 \\
Scam                  & 0.855 \\
Social Isolation      & 0.719
\\\bottomrule
\end{tabular}
\label{tab:auc}
\end{table}

\subsection{Ethical Concern Categorization Results (RQ1)}

\begin{figure*}
\centering
\includegraphics[width=.8\linewidth]{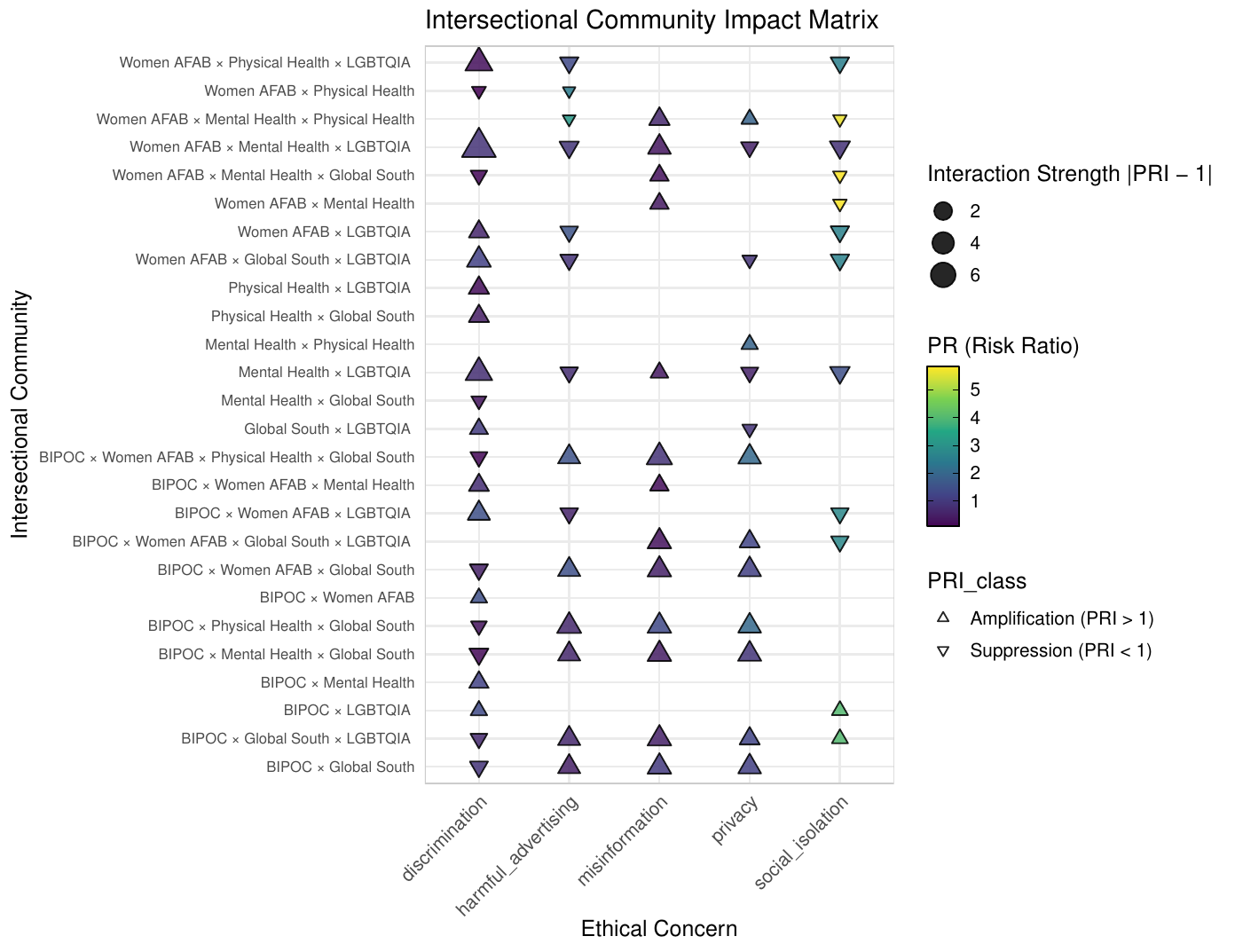}
\caption{This figure illustrates the relative impact of intersectional demographic identities on specific ethical concerns.}
\label{fig:rq1_results}
\end{figure*}

Figure~\ref{fig:rq1_results} summarizes the intersectional prevalence ratio interactions (PRIs) derived from the hierarchical regularized regression models. Model diagnostics indicate moderate discriminatory ability for several ethical concern categories. In particular, the models for \textit{scam} (AUC = 0.855), \textit{misinformation} (AUC = 0.779), \textit{harmful advertising} (AUC = 0.760), \textit{social isolation} (AUC = 0.719), privacy (AUC = 0.714),  and \textit{discrimination} (AUC = 0.700) achieved acceptable levels of predictive discrimination. Other concern categories showed weaker performance (AUC range: 0.640--0.695), indicating that the intersectional patterns for these outcomes should be interpreted cautiously. 

Across concern categories, several intersectional identity combinations exhibited amplification effects (PRI>1), meaning the joint prevalence exceeded what would be expected from the additive contribution of the individual identities. Conversely, suppression effects (PRI<1) indicated that the joint occurrence of these identities reduced the predicted prevalence relative to additive expectations. Scam was the only category for which no PRI values exceeded the corresponding MID values; consequently, it was omitted from Figure~\ref{fig:rq1_results} and the subsequent results section.

\subsubsection{Misinformation}
Misinformation showed consistent amplification across all intersectional communities, with the strongest effects for \textit{BIPOC $\times$ Women AFAB $\times$ Physical Health $\times$ Global South} (PRI = 3.44) and \textit{BIPOC $\times$ Mental Health $\times$ Global South} (PRI = 2.84). No suppression effects were observed for any community. The uniformity of amplification across all intersections distinguishes misinformation from other concern categories, suggesting that marginalized communities broadly perceive elevated misinformation risk regardless of specific identity composition.

\subsubsection{Harmful Advertising}
Amplification for harmful advertising was driven by BIPOC and Global South intersections, most notably \textit{BIPOC $\times$ Physical Health $\times$ Global South} (PRI = 2.98), \textit{BIPOC $\times$ Global South} (PRI = 2.42), and \textit{BIPOC $\times$ Mental Health $\times$ Global South} (PRI = 2.42). Suppression was concentrated among Women/AFAB and LGBTQIA-inclusive intersections, particularly \textit{Women/AFAB $\times$ Mental Health $\times$ LGBTQIA} (PRI = 0.19) and \textit{Women/AFAB $\times$ Physical Health $\times$ LGBTQIA} (PRI = 0.34). This pattern mirrors privacy and discrimination, with Global South intersections amplifying concern expression while LGBTQIA-inclusive intersections suppress it.

\subsubsection{Social Isolation}
Social isolation was predominantly suppressed across intersectional communities, with the strongest suppression for \textit{Women AFAB $\times$ Mental Health $\times$ LGBTQIA} (PRI = 0.03) and \textit{Women AFAB $\times$ Global South $\times$ LGBTQIA} (PRI = 0.34). The only amplification effects were for \textit{BIPOC $\times$ LGBTQIA} (PRI = 1.36) and \textit{BIPOC $\times$ Global South $\times$ LGBTQIA} (PRI = 1.36), both modest in magnitude. The near-universal suppression pattern suggests intersectional communities discuss social isolation at lower rates than predicted by their constituent identities alone.

\subsubsection{Privacy}
Amplification for privacy concerns was driven by BIPOC and Global South intersections, most notably \textit{BIPOC $\times$ Mental Health $\times$ Global South} (PRI = 2.69), \textit{BIPOC $\times$ Global South} (PRI = 2.62), and \textit{BIPOC $\times$ Physical Health $\times$ Global South} (PRI = 2.61). Suppression was observed among LGBTQIA-inclusive intersections, particularly \textit{Women AFAB $\times$ Mental Health $\times$ LGBTQIA} (PRI = 0.51) and \textit{Mental Health $\times$ LGBTQIA} (PRI = 0.51). This amplification/suppression split across Global South and LGBTQIA identity axes mirrors the pattern observed for discrimination and harmful advertising.

\subsubsection{Discrimination}
Amplification was most pronounced for \textit{Women AFAB $\times$ Mental Health $\times$ LGBTQIA} (PRI = 7.37) and \textit{Women AFAB $\times$ Physical Health $\times$ LGBTQIA} (PRI = 4.05), with additional amplification in \textit{Mental Health $\times$ LGBTQIA} (PRI = 3.91) and \textit{Women AFAB $\times$ Global South $\times$ LGBTQIA} (PRI = 2.96). Suppression was concentrated among Global South and health-related intersections, particularly \textit{BIPOC $\times$ Mental Health $\times$ Global South} (PRI = 0.24) and \textit{BIPOC $\times$ Women AFAB $\times$ Physical Health $\times$ Global South} (PRI = 0.55). The consistent pattern of LGBTQIA-inclusive intersections driving amplification while Global South intersections drive suppression suggests divergent discrimination concern expression across these identity axes.

Overall, these findings demonstrate that ethical concerns are not uniformly distributed across demographic groups. Instead, the intersection of multiple identities can either amplify or attenuate concern prevalence, highlighting the importance of intersectional analysis when examining how communities experience potential harms in digital systems.

\subsection{Time Series Analysis Setup (RQ2)}
To answer (RQ2), how ethical concerns are expressed over time by intersectional communities, we applied two complementary forecasting models to monthly proportions of ethical concern posts: ARIMA~\citep{hyndman23forecast,hyndman08forecast}, a standard approach for modeling autocorrelation structure in time series data, and Prophet~\citep{Taylor18prophet}, which was designed for social media-style data and accommodates non-stationary trends, irregular seasonality, and sparse or missing observations common in niche online communities. The dataset consists of Reddit posts aggregated into monthly counts from January 2015 to June 2023. We chose to start the time-series model from 2015 as this period represents the point where the volume reached approximately 50 total posts per month, providing a stable baseline for frequency analysis in accordance with recommended practices for longitudinal data requirements~\citep{hecht2021sample}. We calculated the frequency of reported concerns by normalizing the count of relevant posts against the total post volume.

\subsubsection{World Event Ranking}
\label{sec:worldevent}
We compiled significant news events from the Western hemisphere, aligned with the geographic focus of app rankings and the predominant demographic of Reddit users, starting from the year 2015. To ensure a balanced selection and mitigate potential biases, we consulted multiple reputable news platforms to identify the most salient stories for each year. Notably, CBS's The Year in Review~\citep{year_in_review} provided a particularly comprehensive summary of annually prominent events.

Our selection emphasized news stories with potential societal implications for software use, such as the MeToo movement and the COVID-19 pandemic. The resulting dataset comprised 65 events, available in our replication package. These events were subsequently evaluated by a diverse panel of nine researchers, selected to represent variability in gender, age, ethnicity, professional seniority, Global South/Global North living experiences, and expertise in ethics and software. Each researcher independently rated the relevance of the events to ethical concerns in software applications on a scale from -2 (irrelevant) to +2 (highly relevant). The aggregated scores, calculated as the sum of individual ratings, ranged from -14 to 16, with a median of 2, a mean of 1.292, and a third quartile of 8. To refine the dataset for analysis, we limited events to those with a score above 8 (> third quartile), yielding a final collection of fourteen events.

These fourteen events, which included milestones like the Cambridge Analytica scandal, the COVID-19 pandemic, and Elon Musk’s acquisition of Twitter, were integrated into our models as binary exogenous variables. An Augmented Dickey-Fuller (ADF) test confirmed stationarity (Dickey-Fuller = -6.6884, Lag order = 4, $p = 0.01$), allowing us to proceed with an $ARIMA(0,1,1)$ model with drift. The model fit was validated by a Ljung-Box test on the SARIMAX residuals ($Q* = 15.261, p = 0.08401$), which suggested that the model successfully captured the information in the data without leaving significant autocorrelated noise. We applied a Bonferroni correction to p-values to account for multiple hypothesis testing across these events. Additionally, a Prophet model was used to decompose the time series into a growth trend, yearly seasonality, and holiday effects, with uncertainty intervals of 85\% and 95\% used to identify statistical anomalies.

\subsection{Time Series Analysis Results (RQ2)} 
The final model included a single integrated moving average term (IMA(1,1)) that was large and highly significant ($\beta = -0.79, SE = 0.06, p < .001$), consistent with mean-reverting dynamics in which months with elevated EC expression tend to be followed by a return toward baseline. This pattern suggests that EC activity reflects transient fluctuation rather than sustained directional shifts.
None of the 15 event regressors survived Bonferroni correction (all corrected $ps \geq 1.00$, with the 2016 U.S. Presidential Election as the only term approaching conventional significance prior to correction: $\beta = -0.08, SE = 0.03, p = .008, p_{bonf} = .130$). The negative direction of this coefficient is noteworthy: rather than an increase in EC expression around the election period, the model estimates a slight decrease, which may reflect a temporary displacement of community discourse toward electoral concerns and away from software-specific critique. All remaining events, including GDPR enforcement, the Cambridge Analytica scandal, BLM protests, the Trump platform ban, Frances Haugen's whistleblowing, and the launch of ChatGPT,  showed no statistically reliable association with EC frequency (all uncorrected $ps > .14$). These null findings should be interpreted in light of the event coding scheme: each event was encoded as a single-month impulse, which may underestimate effects that unfolded gradually or with a lag. 

Complementary decomposition of the series using Prophet revealed three additional structural features. First, a modest upward trend in EC frequency across the study period (approximately 0.15 to 0.20), suggesting slow but steady growth in EC expression independent of any specific event. Second, a consistent mid-year seasonal peak, with EC expression rising from approximately April through July before declining in the second half of the year, a pattern likely reflecting broader seasonal rhythms in Reddit engagement rather than any software-specific dynamic. Third, US calendar holidays exerted a small suppressive effect on EC frequency ($range: -0.008 - 0.000$), confirming that holiday periods produce minor reductions in posting activity but do not meaningfully confound the trend or event estimates.

Taken together, these results suggest that EC expression in intersectional communities follows a relatively stable temporal trajectory shaped primarily by gradual trend growth and seasonal rhythms. Real-world platform events do not appear to produce detectable, sustained shifts in EC prevalence at the monthly level, a finding that may reflect the single-month impulse coding scheme's limited sensitivity  to effects that unfold gradually or with a lag.

\begin{figure*}
\centering
\includegraphics[width=.85\textwidth]{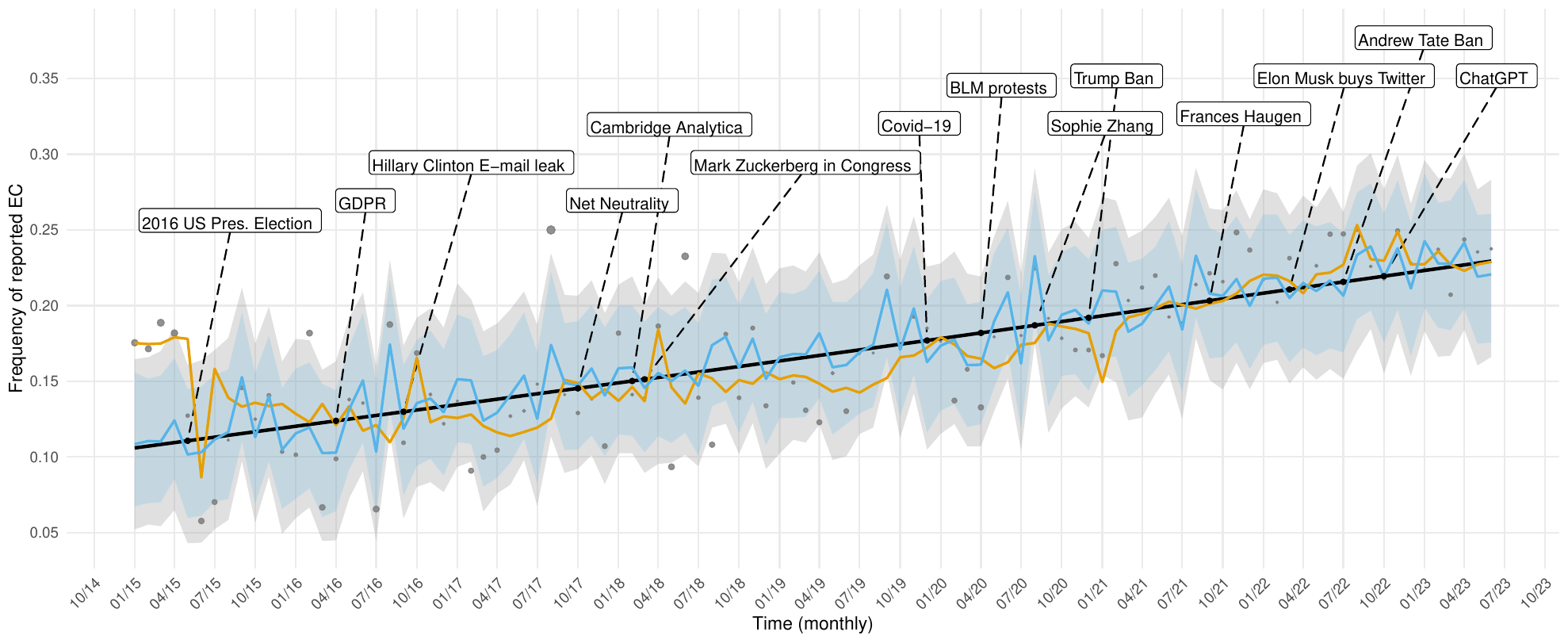}
\caption{Monthly ethical concern frequency (dots), forecasts of ARIMA model (orange line), forecast of Prophet model (blue line), its .95 and .85 CI (blue and grey areas) and trend (black line). Dots grow in size according to their distance to the Prophet trend. World events point at the month of their occurrence.}
\includegraphics[width=.75\textwidth]{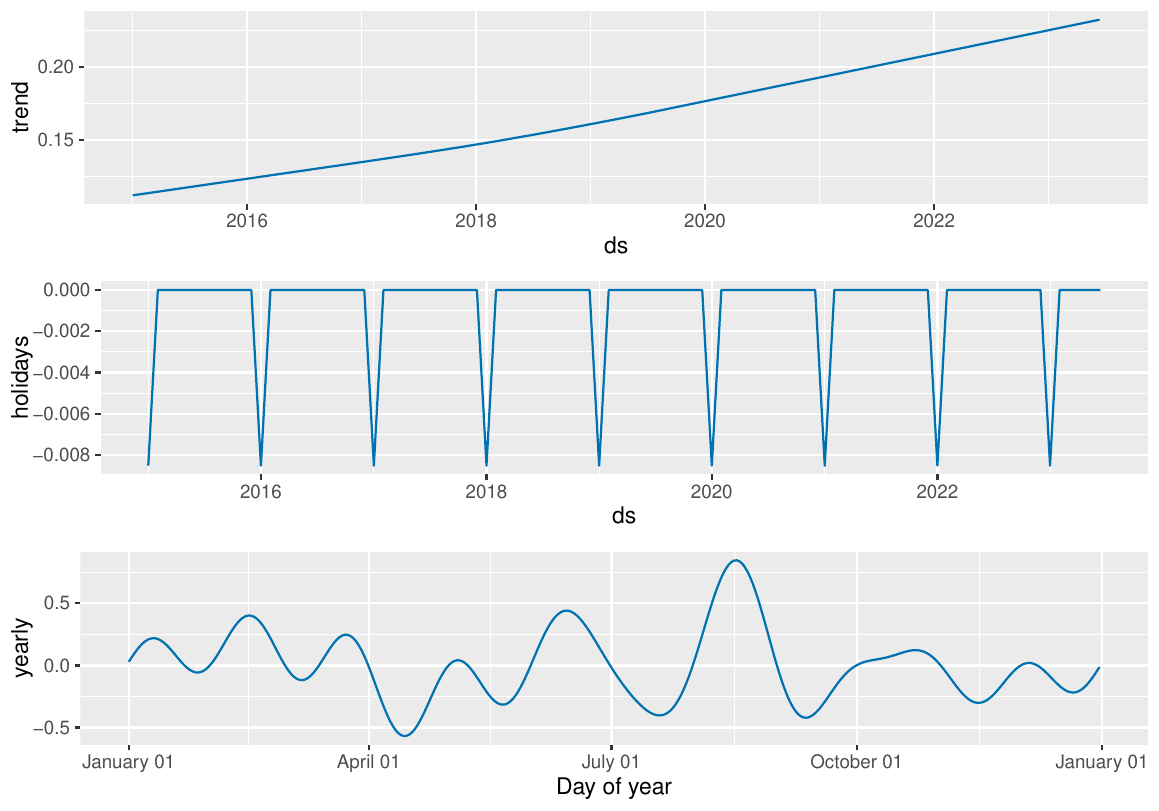}
\caption{Prophet decomposition of monthly ethical concern (EC) frequency, January 2015–December 2023. The trend component (top) shows a gradual increase in normalized EC frequency over the study period ($~0.15 to ~0.20$). The holidays component (middle) captures the estimated effect of US calendar holidays (New Year's Day, Memorial Day, Independence Day, Labor Day, Thanksgiving, and Christmas), which exert a small suppressive effect on EC posting activity (range: $-0.008$ to $0.000$). The yearly component (bottom) reflects consistent seasonal variation, with EC expression peaking in mid-year (approximately April–July) and declining in the second half of the year. Shaded bands represent 95\% uncertainty intervals.}
\label{fig:worldevent_ts}
\end{figure*}


\subsection{Ethical Concerns Prioritization Setup (RQ3)}
To identify which ethical concerns warrant the most urgent attention from the perspective of intersectional users (RQ3), we developed a prioritization framework that scores posts based on multiple factors beyond frequency. We exclude volume because some ethical concerns, although infrequent, possess the potential for severe consequences, particularly in cases relating to violence. We first grouped posts into priority themes and then calculated post-level priority scores by combining indicators of thematic severity (entropy), recency, engagement, and sentiment. These scores were aggregated to estimate the overall priority of each ethical concern category. We validated this method using a user study with 102 participants who rated a stratified sample of posts on perceived importance. The model’s performance was evaluated using precision@k and recall@k, optimized through grid search and cross-validation to balance ranking accuracy with coverage. See Figure 6 for an overview. 

\begin{figure*}
\centering
\includegraphics[width=.65\textwidth]{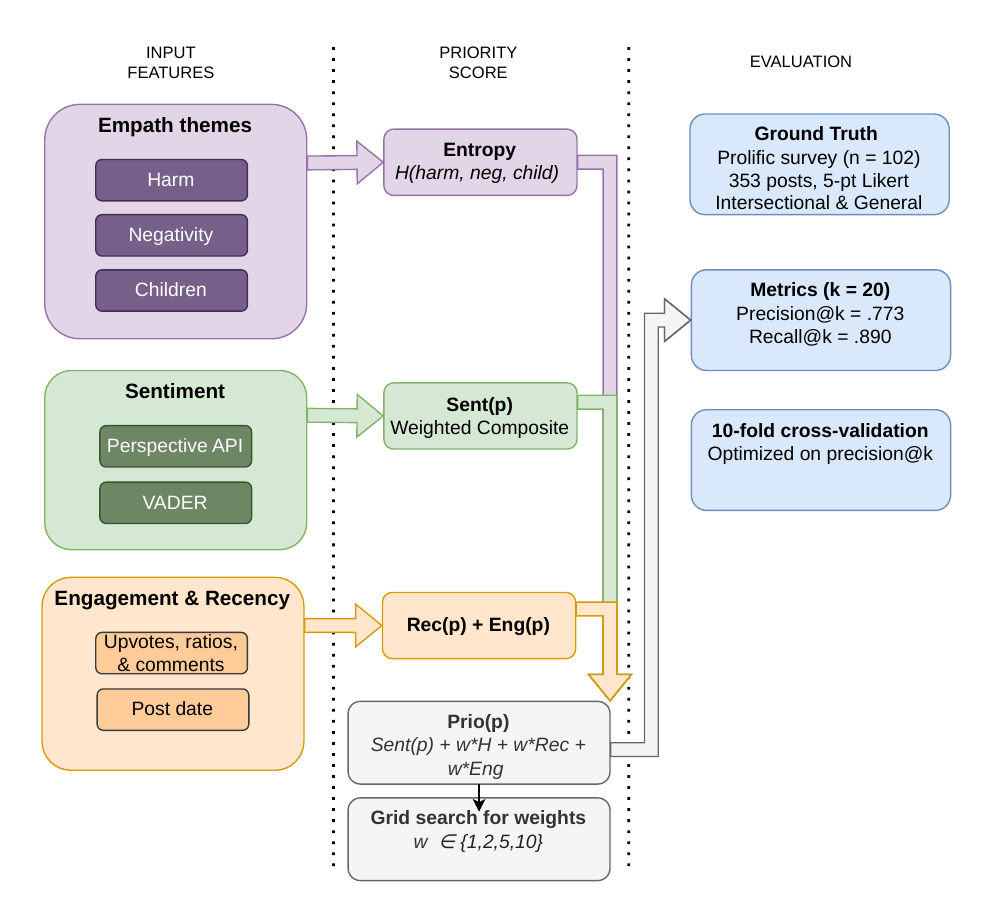}
\caption{Schematic overview of the ethical concern prioritization framework. Post-level features, Empath-derived priority themes (harm, negativity, children), a weighted sentiment composite (Perspective API and VADER), and engagement and recency indicators,  are combined into a post-level priority score Prio(p), with weights optimized via grid search. The scoring system is evaluated against ground truth ratings from a Prolific survey (n=102) using precision@k and recall@k (k=20) with 10-fold cross-validation.}
\label{fig:schematic}
\end{figure*}

\subsubsection{Priority Theme Scoring} 
 While we have already assigned ethical concern categories to our posts, there is variation in the urgency of these ethical concerns. For example, consider the two following ethical concerns that were assigned the same category:

\begin{quote}
    
(1) \textbf{Cyberbullying} (low priority): \textit{``I'm kinda down about something I know is pretty silly in the grand scheme of things. I called out a Facebook group admin for being all passive-aggressive with group members."}
\end{quote}
\begin{quote}
(2) \textbf{Cyberbullying} (high priority): \textit{``She left a suicide note on her Facebook, and the comments were just awful. It's getting really bad now."}
\end{quote}

While these two posts may have the same ethical concerns category, the urgency of their issue varies widely. The first post describes a minor dispute while, in the second, a user threatens suicide and, after, experiences \textit{more} harassment. To ensure the safety of this user and others like her, we identify critical topic terms within our posts.  

To score our posts thematically, we use Empath~\citep{fast2016empath}, a tool for analyzing text across over 200 pre-validated topics. Out of the box, this tool is trained on more than 1.8 billion words of modern fiction. Empath was chosen because its lexical categories were validated against human raters and span a broad range of psychosocial and behavioral dimensions relevant to ethical discourse, making it well-suited for exploratory thematic analysis of community-generated text. Instead, we use their provided Reddit model to generate more relevant scores. Empath is trained via a deep learning skip-gram network to learn word associations. We utilize 21 of Empath's built-in topics to create three \textit{priority themes}: harm, negativity, and children (see Table~\ref{table:topics}). Cases involving children were prioritized because children represent an especially vulnerable population and are less equipped than adults to navigate ethically critical situations, making their protection an essential focus.

We started with a pre-existing set of 15 topics relating to `hate, violence, discrimination, and negative feelings' from previous work on right wing YouTube channels and children's unsafe conversations on Instagram~\citep{10.1145/3201064.3201081, 10.1145/3491102.3501969}; then, after manually inspecting the complete set of topics, we added the following topics: \textit{medical emergency, injury, death, fear, shame}, and \textit{confusion}. We then sorted these topics into three themes. The \textit{harm} theme's topics relate to bodily harm; we developed this theme to capture online mentions of disturbing content. The \textit{negativity} theme captures negative sentiment but in a more fine-grained manner. Finally, we created a \textit{children} theme (with the topics \textit{children} and \textit{youth}) to prioritize ethical concerns that concern young people. Let \( p \) be a post and \( P \) be the set of all posts. Let \( t \) be a topic and \( T \) be the set of topics within a theme \( p_{theme} \) (harm, negativity, youth). For all posts, we sum all the Empath embedding values for each post's priority theme:

\begin{equation}
p_{theme} = \sum_{t \in T} empath(p, t)
\end{equation}

\subsubsection{Measuring Priority}
As mentioned above, we prioritize based on: \textit{A: Entropy of Critical Themes}, \textit{B: Recency}, \textit{C: Engagement} and \textit{D: Sentiment}. 

\begin{table}
\caption{Critical Empath topics in each priority theme}
\footnotesize
\rowcolors{2}{gray!15}{white}
\begin{tabular}{p{0.27\columnwidth}p{0.63\columnwidth}}
\toprule
\textbf{Priority Themes} & \textbf{Empath Topics} \\\midrule
Harm & medical emergency, pain, violence, death, injury, kill, terrorism \\
Negativity & hate, negative emotion, nervousness, suffering, fear, sadness, shame, confusion, aggression, anger, disgust, rage \\
Children & children, youth\\\bottomrule
\end{tabular}
\label{table:topics}
\end{table}

\textbf{A: Entropy of Critical Themes.}
To calculate the relevance of critical themes, we calculate their \textit{entropy}. This is a common anomaly detection metric used, for example, to find product features that are of utmost interest to customers~\citep{somprasertsri2008maximum,takahira2016entropy,zhang2008entropy}. Let $\Phi (p_{theme})$ be the probability of a given post's priority theme value. The entropy \(H(p_{theme})\) of a theme is defined as: 
\begin{equation}
H(p_{theme}) = - (\Phi(p_{theme}) * log_2 \Phi(p_{theme}))
\label{eq:entropy}
\end{equation}
To generate a total entropy score of a given post, we sum the entropy scores of our three priority themes:

\begin{equation}
Ent_{total}(p) = H(p_{harm}) + H(p_{negativity}) + H(p_{children})
\label{eq:total_entropy}
\end{equation}
\textbf{B: Recency.}
We use \textit{recency} as a measure due to the often urgent nature of ethical concerns. For example, if users post content about harming themselves or others, apps must remove this content and contact proper authorities as quickly as possible. Let $integer(p_{date})$ be the date of the post scaled to an integer value. We define the recency of a post as: 
\begin{equation}
Rec(p) = integer(p_{date})
\label{eq:recency}
\end{equation}

\textbf{C: Engagement.}
The \textit{engagement} of a post refers to the amount of reactions a post receives. We posit that the engagement of a post, and therefore the ethical concern described, likely indicates the level of salience this ethical concern has among the online community. Let \(p_{\text{upvotes}}\) be the number of upvotes a post receives, \(p_{\text{ratio}}\) be the ratio of upvotes to downvotes, and \(p_{\text{comments}}\) be the number of comments a post has. Additionally, let \(P_{\text{upvotes}}\) and \(P_{\text{comments}}\) represent the sets of upvote counts and comment counts across all posts, respectively. We define the engagement of a post as:

\begin{equation}
Eng(p) = \frac{p_{upvotes}}{max(P_{upvotes})} + p_{ratio} + \frac{p_{comments}}{max(P_{comments})}
\label{eq:engagement}
\end{equation}

\textbf{D: Sentiment.}
We employ Jigsaw's Perspective API toxicity detector~\citep{perspectiveAPI} and the widely utilized sentiment analysis tool VADER~\citep{hutto2014vader} to capture fine-grained representations of negative sentiment. Prior research indicates that user complaints expressing ethical concerns are often characterized by heightened negativity~\citep{Tjikhoeri2024}. Therefore, incorporating a detailed analysis of negative sentiment can serve as an effective metric for prioritizing ethical concerns in software reviews. Let $Sent(p)$ denote the sentiment of a post. Each variable from the Perspective API, along with the VADER score—namely $Tox(p)$ (toxicity), $Sev(p)$ (severe toxicity), $Ins(p)$ (insult), $Pro(p)$ (profanity), $Thr(p)$ (threat), $Ide(p)$ (identity threat), and $Vad(p)$ (sentiment score)—is assigned an individual weight before being aggregated to form the overall sentiment score. We weight these variables separately due to their diverse nature and salience within this context. 

\begin{equation}
\begin{split}
Sent(p) = w_a \cdot Nor(Tox(p)) + w_b \cdot Nor(Sev(p))\\ + w_c \cdot Nor(Ins(p)) + w_d \cdot Nor(Pro(p))\\ + w_e \cdot Nor(Thr(p)) + w_f \cdot Ide(p) + w_g \cdot Nor(Vad(p))
\end{split}
\label{eq:sentiment}
\end{equation}

\textbf{Normalization.}
Due to the varying scales of the variables, we normalize each variable from 0 to 1. Let $v$ be one of the input variables. Let $V$ be the set of variables. We define normalization as:
\begin{equation}
Nor(v) = \frac{v - min(V)}{max(V) - min(V)}
\label{eq:normalization}
\end{equation}

\textbf{Priority.}  To prioritize according to aforementioned variables, we use the following formula: 
\begin{equation}
\begin{split}
Prio(p) = Sent(p) + w_h \cdot Nor(Ent_{total}(p))\\ + w_i \cdot Nor(Rec(p)) + w_j \cdot Nor(Eng(p)) 
\end{split}
\label{eq:priority}
\end{equation}

\textbf{Weights.} To determine the optimal variable weights for our prioritization equation, we utilize a parameter grid search, evaluating weights from the set {1, 2, 5, 10} for each parameter.  Let \( w_i \) be a variable that can take any value from the set \(\{1, 2, 5, 10\}\), for each \( i \) belonging to the set \(\{a, b, c, d, e, f, g, h, i, j\}\):

\begin{equation}
w_i \in \{1, 2, 5, 10\} \quad \forall i \in \{a,b,c,d,e,f,g,h,i,j\}
\end{equation}
\label{eq:weights}

\subsubsection{Evaluation}
\label{sec:eval}
To evaluate the proposed methodology, it was imperative to establish a reliable ground truth. In alignment with a user-centered approach, we identified this ground truth as the subjective opinions of users. Consequently, we conducted a survey involving 102 participants sourced from the Prolific platform\footnote{https://www.prolific.com/}. The participant pool was bifurcated into two distinct groups: an intersectional group (n=51) and a general population group (n=51), ensuring at least half of the participant pool was from an intersectional background. Each population was surveyed separately to ensure each question had responses from both populations. Both groups comprised individuals proficient in English. Gender balance was maintained across both groups, ensuring a 50:50 male-to-female ratio. We analyze sex rather than gender because Prolific collects sex based on legal documentation, which is the demographic variable available through their platform \footnote{\url{https://researcher-help.prolific.com/en/article/b2943f}}. We acknowledge that this may not align with participants' self-identified gender, particularly for transgender and non-binary individuals for whom legal documentation may not reflect their gender identity. Participants were encouraged from all available countries.

To construct an intersectionally diverse participant group, we implemented specific criteria targeting membership within LGBTQ+ and BIPOC communities. We selected LGBTQ+ criteria to include a broad spectrum of sexual orientations and gender identities. The inclusion of BIPOC participants was strategically aimed to represent both individuals from the Global South and other historically marginalized racial and ethnic groups. Due to the ambiguous nature of socioeconomic status indicators provided by Prolific’s screening tools~\footnote{Q:Where would you put yourself on the socioeconomic ladder? A:[1-10]} and to avoid undue risk to potentially vulnerable groups, we excluded socioeconomic status, mental health and physical health from our criteria. This decision allows for future work to explore these demographics.

The composition of our \textit{final} participant pool was predominantly BIPOC, comprising approximately 69\% (69 out of 102 participants) of the sample. Nearly 60\% of participants were from the Global South, including countries such as South Africa, Mexico, Chile, Nigeria, India, Egypt, Venezuela, Somalia, and Eswatini. The participant demographic also included a balanced representation of women and LGBTQIA+ individuals, each constituting 50\% of the total sample (n = 102). However, specific frequencies of lower socioeconomic status or those with mental or physical health challenges were not measured and remain unknown. 

\textbf{Survey Design.}
\label{section:survey_design}
Our survey, containing 353 total questions, is a representative sample of our dataset of 4,231 labeled ethical concerns (margin of error=.05, confidence level=.95). Each question in the survey represented an ethical concern from a generated random sample of 353  Reddit posts expressing ethical concerns. The first author imported all these posts into Qualtrics, manually shortened them to focus on the ethical concern, and then paraphrased them to protect the authors' privacy. If a post contained an image or a link necessary to view for context or was unclear, we replaced it with an alternate randomly-sampled post. We performed this step to simplify the task for participants - so they did not have to access an external link. Participants evaluated each ethical concern using a 5-point Likert scale to indicate its importance. Participants answered a balanced subset of 20 questions so that each question would have at least 5 responses (5.6 = (20 \text{ answers} * 102 \text{ participants}) / 353 \text{ total questions}). The responses were then averaged to produce an overall ground truth score for each question, facilitating a comprehensive evaluation of ethical concerns within the dataset. Our study aligned with the ethics review procedure at our university, allowing us to proceed without needing a full review from BETHCIE, the Faculty of Science ethics committee at Vrije Universiteit Amsterdam.\footnote{\url{https://vuamsterdam.eu.qualtrics.com/jfe/form/SV_9tBjPqFq6bxv2Sx}} We obtained informed consent from all participants. The full results of the survey are included in the replication package.

\textbf{Metrics.}
\label{sec:weights}
In this evaluation, we used two key metrics to assess the effectiveness of our scoring methodology: \textit{precision@k} and \textit{recall@k}. \textit{Precision} measures the proportion of relevant items, indicating how many of the items are truly relevant. \textit{Recall}, on the other hand, evaluates the proportion of identified relevant items relative to the total number of relevant items available. We use both of these metrics ` at k as these metrics provide insight into our model's ability to capture a comprehensive set of relevant items among its top k recommendations. Our metrics were validated using 10-fold cross validation; this process prevents overfitting by training and testing the set on all data.

We use a grid search to determine an optimal weighting scheme (see Equation~\ref{eq:weights}). The grid search is optimized based on precision@k. We prioritize precision over recall as our optimization metric due to the risk of overfitting in our small dataset, which could result in an artificially high recall. Given the critical nature of ethical concerns in software engineering, it is imperative to prioritize the most significant user concerns at the top of the list, thereby minimizing the occurrence of false positives that could otherwise dilute the focus on genuine issues.

We set the threshold for relevance to 4, aligning with our Likert rating scale where 5 represents ``extremely important," 4 represents ``very important," 3 represents ``moderately important," 2 represents ``slightly important," and 1 represents ``not important." This threshold ensures that only items rated as very important or higher are considered relevant. We chose the value of k to be 20, allowing us to focus on the top 20 recommendations. Together, these metrics and threshold settings offer a balanced perspective on both the accuracy and completeness of our model's recommendations, ensuring that our scoring methodology not only ranks relevant items highly but also captures a wide range of them.

\subsection{Ethical Concerns Prioritization Results (RQ3)}
\begin{figure}
   \centering
 \includegraphics[width=\linewidth]{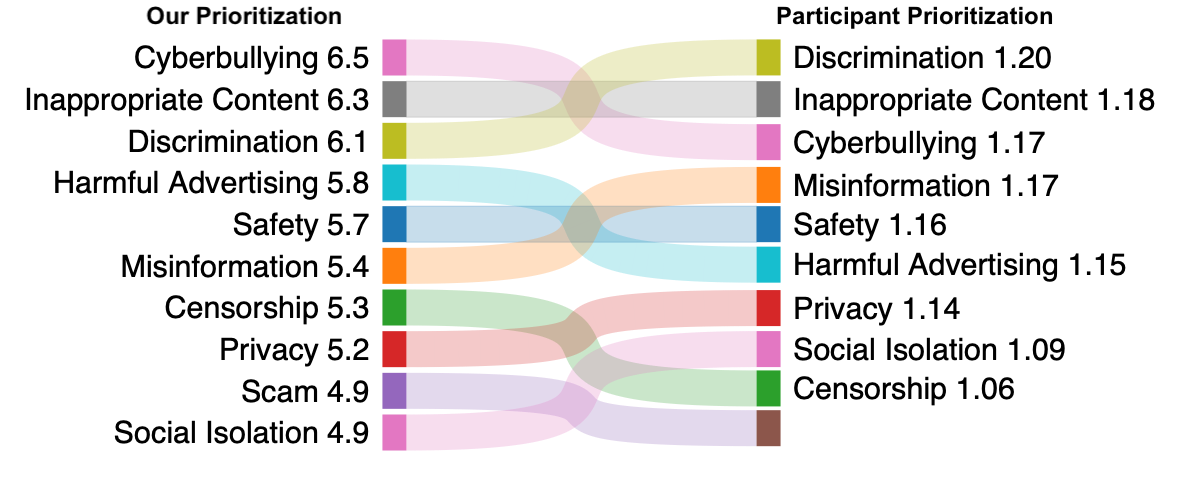}
   \caption{This Sankey diagram illustrates the alignment between our optimized prioritization framework (left) and the average rankings provided by survey participants (right). The algorithmic scores are derived from a weighted sentiment and priority model (Prio(p)), which utilizes a grid-searched weighting scheme. Participant values represent the mean importance rating on a 1–5 scale. The flow lines highlight shifts in priority, such as the increased relative importance of "Discrimination" and "Misinformation" among participants compared to the algorithmic output.} 
   \label{fig:heatmap_priority}
\end{figure}

We find that our prioritization system has a precision and recall (k=20) of .773 and .89, respectively. To calculate the highest priority ethical concerns, we average the priority score of all posts for that concern. The ethical concerns with the highest priority are cyberbullying, inappropriate content, and discrimination; those with the lowest priority are privacy, scam, and social isolation.
Figure~\ref{fig:heatmap_priority} shows the prioritized ethical concerns from highest to lowest priority scores. As illustrated in Figure~\ref{fig:heatmap_priority}, the algorithmic and participant priorities generally align (Spearman's $\rho = 0.778$ (strong), $p = 0.0135$), though participants elevate some concerns (discrimination, misinformation) higher and some lower (censorship) than the model.

We summarise the content of the top ten priority posts of the overall dataset and then examine the ten highest priority user concerns for each ethical concern category. 

\subsubsection{Weights, Recall and Precision}
Our prioritization system has a precision (k=20) of .773 and recall (k=20) of .89. These recall and precision values result from our optimized weighting scheme, which sets recency to 2, entropy and engagement to 5, identity threat to 10, and the rest of the parameters to 1. As a comparison, when all weights are set to one, precision is .25 and recall is 1. A random priority scoring generates a precision of .1 and a recall of 1.

\subsubsection{Top Ten Priority Posts}
Seven out of ten of the top priority posts regard users getting death threats, being told to commit suicide, or discussing the desire to commit suicide. The top post details a young kid being told to commit suicide because of a YouTube video he posted about neo-pronouns. Another top post, from a rape survivor, details hate online they received for discussing pro-choice opinions online. The other two posts detail trauma and how some online content triggered memories of this trauma.

\subsubsection{Priority by Ethical Concern}
\label{sec:results_priority}
We report the specific user concerns for the top 5 ethical concerns (see Figure~\ref{fig:heatmap_priority} by summarizing the top ten concerns from users as follows. 

\textbf{1. Cyberbullying}
Multiple users reported receiving death threats and other violent threats when discussing trans issues online or coming out as transgender. One user recounted that their father sent them transphobic videos on Facebook after they changed their name on the platform, leading to suicidal ideation. Additional posts detailed instances where users' friends threatened suicide, the pervasive presence of transphobia online contributing to suicidal thoughts, and a trans user being told they were `lucky to be raped' because they are unattractive. Another account described a Twitter thread where a 17 year old trans guy was exposed as a groomer, causing him to stop attending school.

\textbf{2. Inappropriate Content}
In this section, we exclude the top two ethical concerns, as they were previously recounted in \textit{Top Ten Priority Posts}. The otherwise top post details an instance where a black woman is killed by domestic violence and her killer publicly jokes about it on Facebook. Subsequently, some reports highlight conflicting viewpoints regarding young transgender individuals on TikTok; one poster describes anger at young trans people on TikTok and the other details frustration with this anger at children. Next, we see a post critiquing the superficial efforts of social media activism aimed at addressing rape. An additional report discusses the particularly insidious nature of TERFism\footnote{trans-exclusionary radical feminism} propagated by gay men on Twitter. After, a transgender user shares their withdrawal from Facebook, driven by the overwhelming presence of transphobic commentary on news articles. Another alarming post discusses the defense of a song on Facebook that glorifies the rape of young Black girls. Lastly, a user reports that the search results for "transgender" on YouTube reveals a predominance of negative and transphobic content.

\textbf{3. Discrimination}
Misgendering, particularly via non-binary-inclusive terms like using ``transmasc" instead of ``transman," is reported by one user. A Black woman user describes receiving backlash from black men on Twitter for posting about Black womens' experiences of sexual assault and domestic violence on the site. Transphobia is reportedly worsening, exemplified by an incident where a woman with breast cancer and a double mastectomy receives transphobic hate. One user expresses regret about transitioning due to societal focus on trans issues. Additionally, a trans user fears ridicule and harassment when posting selfies, with concerns about images being cross-posted on Twitter by transphobic individuals. Another user laments on how TERFs deadname\footnote{use of a trans person's birth name} a murdered trans woman on Twitter. Another trans individual reports being told to remain silent about transphobia because other groups, such as Black and Asian communities, suffer more. Users also report the mass bullying of a transmedicalist YouTuber and the fetishization of gay relationships within the TikTok trans community.

\textbf{4. Harmful Advertising}
Several users reported frustration with the receipt of transphobic and anti-abortion advertisements on YouTube. Despite attempts to modify ad preferences—specifically requests not to see such ads—these users noted that the undesired content continued to appear, indicating a potential failure in the platform's ad targeting and preference management systems. Furthermore, some users who had recently experienced miscarriages detail being subjected to targeted baby product advertisements. This mismatch between user experiences and ad targeting led to reports of significant emotional trauma. Additionally, one user detailed their experience with targeted advertisements for questionable dietary supplements, which were promoted during that users' menstrual periods. Finally, one transman reports feeling `like a bad woman' after receiving YouTube ads targeted at weight loss for ciswomen.

\textbf{5. Safety.}
The highest priority post is a report of a user's girlfriend threatening suicide on Discord and then losing contact. Other posts within this category disclose feeling upset after seeing TikTok videos that blame pregnant people's stress levels for miscarriage or online pregnancy announcements. Another post details the harassment of a trans school teacher on a town Facebook page, leading him to fear for his safety in public. Finally, one poster discusses their gender dysphoria's fluctuations and questions social media's role in their trans identity.  

 In summary, the prioritization system achieved a high precision (0.773) and recall (0.89) with optimized weighting, effectively ranking cyberbullying, inappropriate content, and discrimination as the top ethical concerns, while privacy, scams, and social isolation ranked lowest. Key findings highlight that the highest-priority posts often involve severe emotional harm, such as death threats, transphobia, or triggering online content, particularly targeting marginalized groups like transgender individuals and Black women. Additionally, harmful advertising and safety concerns further reveal systemic failures in content moderation and platform algorithms, exacerbating emotional distress for users in vulnerable situations.

\section{Discussion and Conclusion}
 We find that \textbf{(RQ1)} ethical concern expression is amplified for some communities and suppressed for others; \textbf{(RQ2)} these concerns are increasing and irresponsive to real-world factors; and \textbf{(RQ3)} critical user complaints regarding social media often include descriptions of trauma. Our findings are directly relevant to practitioners who can use our methodology and results to make their software more ethical for intersectional communities. 

\textbf{(RQ1)} The findings from this study carry practical implications for how developers, platform operators, and external auditors respond to ethically significant user feedback from intersectional communities. 

Our results reveal that ethical concern expression is not additive across identities. The amplification of misinformation concerns for BIPOC × Women AFAB × Physical Health × Global South communities (PRI = 3.44) alongside the near-total suppression of discrimination concerns for BIPOC × Mental Health × Global South (PRI = 0.24), despite their strong amplification for Women AFAB × Mental Health × LGBTQIA+ (PRI = 7.37),  illustrates that aggregated feedback obscures divergent experiences only visible at the intersection level. This has direct consequences for update prioritization: a team monitoring privacy concerns in aggregate would see the 2.6x amplification for BIPOC × Global South users (PRI = 2.62) and the concurrent suppression among Mental Health × LGBTQIA+ users (PRI = 0.51) cancel each other out, producing a misleadingly moderate signal.

Gokgoz et al.~\citep{gokgoz2025if} demonstrate that the rewards of responding to user feedback and the penalties of ignoring it can be substantial, with effects contingent on both feedback topic and update timing. Standard developer feedback pipelines are not equipped to surface these patterns. Integrating the hierarchical regularized classification pipeline described here into social listening tools (like Sprinklr, used by Microsoft~\citep{sprinklrmicrosoft2025}) or internal systems (see Facebook~\citep{analyticsfacebook2025}) would allow organizations to move beyond aggregate sentiment tracking, for instance, detecting that some ethical concerns are dramatically amplified for specific intersectional communities while suppressed for others, enabling targeted rather than broadcast responses. This limitation is not addressed by existing platform-level solutions. Both the Apple App Store and Google Play Store now offer AI-generated review summaries for developers~\citep{applellm2024, kearnsnew2025}, but these tools aggregate across all users by default, with no mechanism for detecting whether specific ethical concerns are amplified or suppressed for particular intersectional communities. A summary that surfaces "privacy" as a top concern tells a developer nothing about whether that concern is driven 2.6x above baseline by BIPOC × Global South users while simultaneously suppressed among LGBTQIA+-inclusive communities, preventing developers from serving all users equally. 

Our classification methodology also supports longitudinal external auditing that treats app updates as observable responses to user feedback. Gokgoz et al.'s analysis~\citep{gokgoz2025if} of over one million reviews matched against 3,255 app updates demonstrates that developers' choices about which feedback to address have measurable downstream effects on ratings. Extended to intersectional ethical concern data, this suggests a tractable auditing protocol: classify ethical concerns by community and concern type, match their temporal distribution against release notes, and assess whether amplified communities, such as BIPOC × Global South users raising harmful advertising concerns at nearly 3x expected rates, received commensurate developer attention. This constitutes a form of aggregate discrimination audit: the pattern of whose concerns are systematically deprioritized only becomes visible when thousands of posts are disaggregated by identity intersection.

Our study represents the classification and ethical concern identification step in this pipeline. Future work should operationalize the downstream auditing step, matching classified concerns against changelogs to assess whether intersectional amplification predicts developer inaction, and whether inaction predicts subsequent user dissatisfaction. Where Gokgoz et al.~\citep{gokgoz2025if} show that mismatching user feedback with development decisions carries measurable rating costs, an intersectional extension asks whose feedback is systematically mismatched, and at what cost to underrepresented communities.

\textbf{(RQ2)} The time series results suggest that ethical concern expression in intersectional communities follows a stable, slowly growing trajectory, one shaped more by gradual trend and seasonal rhythm than by discrete platform events. The mean-reverting dynamics captured by the IMA(1,1) term indicate that spikes in ethical concern expression are transient rather than indicative of sustained directional shifts. The modest upward trend identified by Prophet (approximately 0.15 to 0.20 across the study period) nevertheless implies that intersectional communities are, over time, expressing ethical concerns about software at increasing rates,  a signal that developer and auditor attention to this feedback should scale accordingly.

From a practical standpoint, the temporal pipeline introduced here represents an open-source complement to existing proprietary developer analytics. Google Play Console's review analytics already allow developers to monitor how user sentiment and feedback topics shift over time~\citep{kearnsnew2025}. The primary limitation of the current temporal implementation is the use of GPT-4.1 mini for EC classification, which introduces cost and reproducibility constraints. As noted in Section~\ref{sec:multilabels}, future iterations of this pipeline will prioritize open-source classifiers, making the full workflow, from intersectional EC classification through temporal modeling, accessible without proprietary API dependence.

The null findings for world events, including the Cambridge Analytica scandal, GDPR enforcement, BLM protests, and the launch of ChatGPT, warrant careful interpretation. Rather than indicating that these events had no effect on community discourse, they may reflect a methodological limitation of single-month impulse coding, which is ill-suited to capturing effects that unfold gradually or with a lag.  The absence of these signals may also reflect the nature of the posts themselves: ethical concerns in this dataset typically describe personal, first-person accounts of intersectional experience, where the mention of an ethical concern is not the main focus (e.g., `I felt really excluded today at work, and it was made worse when my queer post was tagged as inappropriate by Facebook'), so discussion is unlikely to be influenced by specific, tech-related world events. The 2016 U.S. Presidential Election's negative coefficient, though not surviving Bonferroni correction, offers a suggestive illustration: rather than amplifying software-specific EC expression, the election period may have temporarily displaced community attention away from software critique altogether. Future work employing distributed lag models~\citep{gasparrini2010distributed} or change-point detection~\citep{truong2020selective} would be better positioned to capture these dynamics.

\textbf{(RQ3)} Recent research~\citep{li2023unveiling} has demonstrated a need for user feedback tools for prioritization and thematic analysis. As a solution, we propose a novel methodology for identifying high-priority ethical concern feedback from users that practitioners can use to find critical feedback, though it may initially be most accurate within this specific data context and require data manipulation and addition to be generalizable. The novelty of this system lies in the \textit{entropy of critical themes} and fine-grained sentiment variables, which allow us to sort posts by the urgency of the content from a \textit{user perspective}. Our current system categorizes harmful advertising as the fourth highest ethical concern despite its lack of volume; it is the second least frequent concern in our dataset. Prioritization systems typically include volume as a metric for priority; however, we chose to exclude volume as a measure because some ethical concerns may happen infrequently but can have potentially devastating effects. In addition, some intersectional communities' feedback may be lower in volume due to their marginalization; this should not result in lower priority. 

For addressing high-priority concerns of intersectional communities, we find that the majority of issues relate to traumatic content. We find that Scott et al.'s \textit{Trauma-Informed Social Media Design} presents particularly salient solutions~\citep{10.1145/3544548.3581512}. This trauma-informed framework acknowledges `that everyone likely has a trauma history' and considers the multiple levels at which people experience trauma (e.g., developmental, cultural, community). The first relevant design feature, termed `context moderation,' is based on the idea that contemporary content moderation practices often overlook the situational context, such as the author and the target audience. Consider two cases from Section~\ref{sec:results_priority}: a father sending his recently-out trans daughter (who had just updated her Facebook name) anti-trans content on Facebook, and women who have experienced miscarriages receiving targeted ads for baby products. When considering the context of these scenarios, the content becomes more harmful. If top-down automated content moderation techniques (where platforms remove content) accounted for contextual factors, they could more accurately identify critically harmful content. To infer context, platforms could implement a manipulable design feature allowing users to select their preferences for seeing potentially trauma-inducing content~\citep{bradford2019report}.

It is essential to consider, when developing new software features related to content moderation, that preferences vary significantly by identity. For instance, a recent study found that transgender users find exposure to be undesirable~\citep{schoenebeck2021drawing}; however, most modern social media platforms tend to prioritize engagement highly. Several high-priority reports from Section~\ref{sec:results_priority} detail unwanted attention online, such as the unintended virality of a trans user's Tweet describing sexual assault, a TikTok video about coming out as trans or a trans user worrying that their selfie would be mocked \textit{across} platforms. Allowing users to customize audience targeting and thus size, including tools to disallow screenshotting and nonconsensual spread of their images, may help prevent malicious actors from interacting with their content. Furthermore, to mitigate death threats, suicide baiting, and other forms of violent harassment, we suggest platforms take inspiration from Block Party app's Lockout Filters~\footnote{https://www.blockpartyapp.com/how-it-works/}, which trauma-informed design considers crucial for giving users `power and agency to respond to mass harm' \citep{10.1145/3544548.3581512}.

While the survey was primarily used to evaluate the prioritization framework, it also offers important implications for future work on ethical concern modeling. The inclusion of both intersectional and general user populations enables a broader grounding of the prioritization scores in lived user perspectives. Although we did not conduct a full comparative or qualitative analysis of these data, the responses provided a critical reference baseline for evaluating the alignment between automated prioritization and perceived ethical urgency. This methodological step strengthens the validity of our approach and underscores the importance of incorporating diverse user judgments into algorithmic evaluations. Future studies could build on this dataset to explore how different demographic groups weigh ethical harms, revealing disparities in concern perception that may otherwise be obscured. Highlighting the survey's evaluative function thus not only reinforces the integrity of our framework but also points to fertile directions for advancing ethical software research grounded in user experience.

Our findings indicate that social media apps should make significant changes to safeguard the welfare of intersectional communities. Regarding \textit{scams}, app features should allow users to communicate feedback to businesses and fellow consumers easily and clearly to combat fraudulent business practices online. Business accounts should have separate design features from users to be held accountable and easily distinguished. 

Our dataset contains a disproportionate number of posts from communities identifying as Women/AFAB, comprising over 80\% of the total sample. This overrepresentation is a reflection of Reddit's community landscape, where gendered subreddits for marginalized users are particularly active and vocal about ethical concerns. As a result, the ethical concerns about software surfaced in this paper are likely to be influenced by the specific priorities and lived experiences of these communities. For instance, our qualitative analysis highlights concerns related to reproductive privacy and trans identity, issues that are especially salient within Women/AFAB communities.


There is a significant risk in developing automated methods to mine intersectional communities' ethical concerns feedback; this process is potentially \textit{extractive} and \textit{tokenistic}~\citep{costanza2020design}. Our method is potentially \textit{extractive} because we mine the ideas of intersectional communities to be built into marketed products~\citep{costanza2020design}. Currently, practitioners surveil users' app engagement~\citep{li2023unveiling}. They \textit{extract} user behavior as implicit feedback and use it to change products for their fiscal benefit. By contrast, our approach considers users' more \textit{explicit} feedback by taking their voices into account. However, it is important to note that gathering real-time user feedback could also lead to mass surveillance of marginalized groups.

\section{Limitations}
Our observations and conclusions are constrained by several important limitations that must be considered when interpreting or attempting to replicate this work. First, our findings are restricted to the U.S. context: we relied on U.S. top app rankings, analyzed only English-language posts, and focused on Reddit, a platform predominantly used by U.S. residents. Many of the news events we collected also centered on the Western hemisphere, further limiting global applicability. A further limitation of Reddit is the restriction imposed by Reddit’s API, which allows retrieval of only the most recent 1000 posts per subreddit; as a result, our dataset may underrepresent older discussions and long-term trends in intersectional users’ ethical concerns.

 The representativeness of our findings is another important limitation. Individuals who do not engage on Reddit, or who are less likely to publicly share ethical concerns, are not captured in this analysis. As a result, our insights likely reflect a subset of more digitally active or platform-savvy users. To partially mitigate this threat, we conducted a follow-up survey that included both general and intersectional participants, offering a complementary perspective. However, the survey sample, though diverse in some dimensions, remains relatively small and cannot fully compensate for the broader limitations in representativeness.

 While we did not have the resources to verify the identities of all individuals posting within intersectional subreddits, we assume that the majority of contributors are members of the communities the subreddits represent. In cases where posters may not belong to the community, their participation is still likely to center on issues relevant to that group’s lived experience, thus contributing meaningfully to the discourse we aimed to capture.

 Selection bias may also affect both the survey participant pool and the apps included in our analysis. Our selection of apps was based on frequency of mention, which led to a focus on social media platforms. While this choice allowed us to conduct multi-class classification with sufficient data for model training, it limits the generalizability of our results to other software types. Multiclass classification models require adequate data per class to perform reliably, and the small number of posts in other domains made classification infeasible. Although narrowing our scope reduces the breadth of our findings, it enabled a more robust and interpretable analysis in a domain where intersectional users are particularly active and where ethical concerns are most frequently raised.

Our ethical concern taxonomy also has known limitations. Some categories overlap conceptually, such as misinformation and harmful advertising~\citep{10628454}, which may introduce inconsistencies in classification. Additionally, the use of GPT-4 for assigning concern types introduces potential bias due to the model's underlying training data and algorithmic constraints. The fine-tuned GPT-4.1-mini model used to scale labels to the full dataset achieved 76.12\% validation accuracy, suggesting that classification error propagates into the prevalence ratios and glinternet results used to answer RQ1. We mitigate this risk in part by relying on Tjikhoeri et al.'s~\citep{Tjikhoeri2024} externally developed and validated ethical concern definitions, this adds detail and specificity to the prompt which may allow for greater future reproducibility and minimize bias across models. Nonetheless, further research is needed to understand how such biases may have influenced our results. One direction could be to use an `LLM jury' approach, where multiple labels are generated per post and inter-model disagreement is assessed and adjudicated to increase reproducibility and decrease bias~\citep{verga2024replacing, zhao2025language}. To protect user privacy, the first author paraphrased posts, which introduces some subjectivity despite efforts to preserve the original intent.

Temporal sparsity further limits our analysis: posts prior to 2018 were relatively few, restricting early time series insights. Terminological ambiguity also affected results; for example, app names like “McDonald’s” were excluded unless the word “app” was explicitly mentioned, likely undercounting some platforms.

We also acknowledge the role of researcher positionality. Five researchers identify as feminists and engage in social justice work, and three identify as members of an intersectional community. While our shared values shaped the research questions and goals, we aimed to reduce subjectivity during annotation through collaboratively defined guidelines and a focus on author-expressed concerns rather than annotator interpretations. Broader demographic representation among annotators would further strengthen the reliability of this work.

Finally, the prioritization system we developed has its own limitations. The evaluation sample (n=102) was too small for broad generalization, and the use of a 5-point Likert scale limited response nuance. The system’s weighting method is also vulnerable to overfitting due to the limited data size. As such, we present this system as an early prototype for identifying critical intersectional concerns rather than a deployable prioritization tool. Additionally, recent Reddit policy changes have rendered many subreddits inaccessible, limiting future data collection and replication efforts.
\section{Declarations}
\subsection{Funding}
No funding was received for this work.
\subsection{Ethical Approval}
We obtained ethical approval from our university and consent from our survey participants, more information regarding this can be found in Section~\ref{section:survey_design}.
\subsection{Author Contributions}
Lauren Olson contributed to the study design, developed the machine learning models and the prioritization system, conducted the survey, performed data annotation and analysis, and drafted the manuscript. Tom P. Humbert contributed to the study design, implemented time series analyses, assisted with data annotation and analysis, and contributed to writing. Ricarda Anna-Lena Fischer contributed to the study design, participated in data annotation and analysis, and contributed to writing the manuscript. Bob Westerveld optimized the prioritization system. Florian Kunneman provided guidance as an advisor. Emitzá Guzmán contributed to the study design, oversaw the research as the main advisor, contributed to writing, and provided critical feedback throughout the study. 
\subsection{Data Availability Statement}
Data is available at \url{www.doi.org/10.6084/m9.figshare.31999482}. 
\subsection{Conflicts of Interest}
We have no conflicts of interest.
\subsection{Clinical Trial Number}
Not applicable. 

\bibliographystyle{spbasic}  
\bibliography{citations}

@incollection{crenshaw2013demarginalizing,
  title={Demarginalizing the intersection of race and sex: A black feminist critique of antidiscrimination doctrine, feminist theory and antiracist politics},
  author={Crenshaw, Kimberl{\'e}},
  booktitle={Feminist legal theories},
  pages={23--51},
  year={2013},
  publisher={Routledge}
}

@incollection{crenshaw2013mapping,
  title={Mapping the margins: Intersectionality, identity politics, and violence against women of color},
  author={Crenshaw, Kimberl{\'e} Williams},
  booktitle={The public nature of private violence},
  pages={93--118},
  year={2013},
  publisher={Routledge}
}

@misc{perspectiveAPI,
  author = {Jigsaw},
  title = {Perspective API},
  year = {2024},
  howpublished = {\url{https://www.perspectiveapi.com}},
  note = {Accessed: 2024-06-20}
}

@inproceedings{hutto2014vader,
  author = {C.J. Hutto and Eric Gilbert},
  title = {{VADER: A Parsimonious Rule-based Model for Sentiment Analysis of Social Media Text}},
  booktitle = {Proceedings of the International AAAI Conference on Web and Social Media},
  year = {2014},
  url = {http://comp.social.gatech.edu/papers/icwsm14.vader.hutto.pdf},
  note = {Accessed: 2024-06-20}
}

@article{bradford2019report,
  title={Report of the Facebook data transparency advisory group},
  author={Bradford, Ben and Grisel, Florian and Meares, Tracey L and Owens, Emily and Pineda, Baron L and Shapiro, Jacob N and Tyler, Tom R and Peterman, Danieli Evans},
  journal={Yale Justice Collaboratory},
  year={2019}
}

@article{schoenebeck2021drawing,
  title={Drawing from justice theories to support targets of online harassment},
  author={Schoenebeck, Sarita and Haimson, Oliver L and Nakamura, Lisa},
  journal={new media \& society},
  volume={23},
  number={5},
  pages={1278--1300},
  year={2021},
  publisher={Sage Publications Sage UK: London, England}
}

@article{verga2024replacing,
  title={Replacing judges with juries: Evaluating llm generations with a panel of diverse models},
  author={Verga, Pat and Hofstatter, Sebastian and Althammer, Sophia and Su, Yixuan and Piktus, Aleksandra and Arkhangorodsky, Arkady and Xu, Minjie and White, Naomi and Lewis, Patrick},
  journal={arXiv preprint arXiv:2404.18796},
  year={2024}
}

@inproceedings{zhao2025language,
  title={Language model council: Democratically benchmarking foundation models on highly subjective tasks},
  author={Zhao, Justin and Plaza-del-Arco, Flor Miriam and Genchel, Benjamin and Curry, Amanda Cercas},
  booktitle={Proceedings of the 2025 Conference of the Nations of the Americas Chapter of the Association for Computational Linguistics: Human Language Technologies (Volume 1: Long Papers)},
  pages={12395--12450},
  year={2025}
}

@inproceedings{10.1145/3544548.3581512,
author = {Scott, Carol F and Marcu, Gabriela and Anderson, Riana Elyse and Newman, Mark W and Schoenebeck, Sarita},
title = {Trauma-Informed Social Media: Towards Solutions for Reducing and Healing Online Harm},
year = {2023},
isbn = {9781450394215},
publisher = {Association for Computing Machinery},
address = {New York, NY, USA},
url = {https://doi.org/10.1145/3544548.3581512},
doi = {10.1145/3544548.3581512},
booktitle = {Proceedings of the 2023 CHI Conference on Human Factors in Computing Systems},
articleno = {341},
numpages = {20},
location = {Hamburg, Germany},
series = {CHI '23}
}

@inproceedings{Olson2023,
author={Olson, Lauren and Guzmán, Emitzá and Kunneman, Florian},
  booktitle={2023 IEEE/ACM 45th International Conference on Software Engineering: Software Engineering in Society (ICSE-SEIS)}, 
  title={Along the Margins: Marginalized Communities’ Ethical Concerns about Social Platforms}, 
  year={2023},
  volume={},
  number={},
  pages={71-82},
  doi={10.1109/ICSE-SEIS58686.2023.00013}}

@article{tizard2022voice,
  title={Voice of the users: an extended study of software feedback engagement},
  author={Tizard, James and Rietz, Tim and Liu, Xuanhui and Blincoe, Kelly},
  journal={Requirements Engineering},
  volume={27},
  number={3},
  pages={293--315},
  year={2022},
  publisher={Springer}
}

@INPROCEEDINGS{10628454,
  author={Kara\c{c}am, \"{O}zge and Humbert, Tom P. and Guzm\'{a}n, Emitz\'{a}},
  booktitle={2024 IEEE 32nd International Requirements Engineering Conference (RE)}, 
  title={Uncovering Patterns in Users' Ethical Concerns About Software}, 
  year={2024},
  volume={},
  number={},
  pages={466-474},
  doi={10.1109/RE59067.2024.00055}}

@article{oakley2016disturbing,
	title        = {Disturbing hegemonic discourse: Nonbinary gender and sexual orientation labeling on Tumblr},
	author       = {Oakley, Abigail},
	year         = 2016,
	journal      = {Social Media+ Society},
	publisher    = {SAGE Publications Sage UK: London, England},
	volume       = 2,
	number       = 3,
	pages        = 2056305116664217
}

@article{li2023unveiling,
  title={Unveiling the Life Cycle of User Feedback: Best Practices from Software Practitioners},
  author={Li, Ze Shi and Arony, Nowshin Nawar and Devathasan, Kezia and Sihag, Manish and Ernst, Neil and Damian, Daniela},
  journal={arXiv preprint arXiv:2309.07345},
  year={2023}
}

@inproceedings{somprasertsri2008maximum,
  title={A maximum entropy model for product feature extraction in online customer reviews},
  author={Somprasertsri, Gamgarn and Lalitrojwong, Pattarachai},
  booktitle={2008 IEEE Conference on Cybernetics and Intelligent Systems},
  pages={575--580},
  year={2008},
  publisher={IEEE},
  address={Chengdu, China}
}

@article{takahira2016entropy,
  title={Entropy rate estimates for natural language—a new extrapolation of compressed large-scale corpora},
  author={Takahira, Ryosuke and Tanaka-Ishii, Kumiko and Debowski, Lukasz},
  journal={Entropy},
  volume={18},
  number={10},
  pages={364},
  year={2016},
  publisher={MDPI}
}

@inproceedings{zhang2008entropy,
  title={An entropy-based model for discovering the usefulness of online product reviews},
  author={Zhang, Richong and Tran, Thomas},
  booktitle={2008 IEEE/WIC/ACM International Conference on Web Intelligence and Intelligent Agent Technology},
  volume={1},
  pages={759--762},
  year={2008},
  publisher={IEEE},
  address={Sydney, Australia}
}

@inproceedings{fast2016empath,
  title={Empath: Understanding topic signals in large-scale text},
  author={Fast, Ethan and Chen, Binbin and Bernstein, Michael S},
  booktitle={Proceedings of the 2016 CHI conference on human factors in computing systems},
  pages={4647--4657},
  year={2016},
  publisher={ACM},
  address={San Jose, CA, USA}
}

@article{Tjikhoeri2024,
  author    = {Tjikhoeri, N. and Olson, L. and Guzm{\'a}n, E.},
  title     = {The best ends by the best means: ethical concerns in app reviews},
  journal   = {Empirical Software Engineering},
  volume    = {29},
  pages     = {138},
  year      = {2024},
  doi       = {10.1007/s10664-024-10463-7},
  url       = {https://doi.org/10.1007/s10664-024-10463-7}
}

@misc{reddit_2022,
	title      = {Reddit users by country 2022},
    publisher          = {World Population Review},
	url          = {https://worldpopulationreview.com/country-rankings/reddit-users-by-country},
    year={2022},
    author={WPR}
}

@misc{Boe_2016, title={Praw-dev/PRAW: PRAW, an acronym for “Python reddit api wrapper”, is a python package that allows for simple access to Reddit’s API.}, url={https://github.com/praw-dev/praw}, journal={GitHub}, author={Boe, Bryce}, year={2016}}

@inproceedings{kim2022designing,
  title={Designing chatbots with black americans with chronic conditions: Overcoming challenges against covid-19},
  author={Kim, Junhan and Muhic, Jana and Robert, Lionel Peter and Park, Sun Young},
  booktitle={Proceedings of the 2022 CHI Conference on Human Factors in Computing Systems},
  pages={1--17},
  year={2022}, 
  publisher={ACM},
  address={New Orleans, LA, USA}
}

@inproceedings{10.1145/3491102.3501969,
author = {Ali, Shiza and Razi, Afsaneh and Kim, Seunghyun and Alsoubai, Ashwaq and Gracie, Joshua and De Choudhury, Munmun and Wisniewski, Pamela J. and Stringhini, Gianluca},
title = {Understanding the Digital Lives of Youth: Analyzing Media Shared within Safe Versus Unsafe Private Conversations on Instagram},
year = {2022},
isbn = {9781450391573},
publisher = {Association for Computing Machinery},
address = {New York, NY, USA},
url = {https://doi.org/10.1145/3491102.3501969},
doi = {10.1145/3491102.3501969},
booktitle = {Proceedings of the 2022 CHI Conference on Human Factors in Computing Systems},
articleno = {148},
numpages = {14},
location = {New Orleans, LA, USA},
series = {CHI '22}
}

@inproceedings{10.1145/3201064.3201081,
author = {Ottoni, Raphael and Cunha, Evandro and Magno, Gabriel and Bernardina, Pedro and Meira Jr., Wagner and Almeida, Virg\'{\i}lio},
title = {Analyzing Right-Wing YouTube Channels: Hate, Violence and Discrimination},
year = {2018},
isbn = {9781450355636},
publisher = {Association for Computing Machinery},
address = {New York, NY, USA},
url = {https://doi.org/10.1145/3201064.3201081},
doi = {10.1145/3201064.3201081},
booktitle = {Proceedings of the 10th ACM Conference on Web Science},
pages = {323–332},
numpages = {10},
location = {Amsterdam, Netherlands},
series = {WebSci '18}
}

@book{costanza2020design,
  title={Design justice: Community-led practices to build the worlds we need},
  author={Costanza-Chock, Sasha},
  year={2020},
  publisher={The MIT Press}
}

@article{hedditch2023design,
  title={Design Justice in Practice: Community-led Design of an Online Maker Space for Refugee and Migrant Women},
  author={Hedditch, Sonali and Vyas, Dhaval},
  journal={Proceedings of the ACM on Human-Computer Interaction},
  volume={7},
  number={GROUP},
  pages={1--39},
  year={2023},
  publisher={ACM New York, NY, USA}
}

@article{kumar2020taking,
  title={Taking the long, holistic, and intersectional view to women’s wellbeing},
  author={Kumar, Neha and Karusala, Naveena and Ismail, Azra and Tuli, Anupriya},
  journal={ACM Transactions on Computer-Human Interaction (TOCHI)},
  volume={27},
  number={4},
  pages={1--32},
  year={2020},
  publisher={ACM New York, NY, USA}
}

@inproceedings{wong2018designing,
  title={Designing for intersections},
  author={Wong-Villacres, Marisol and Kumar, Arkadeep and Vishwanath, Aditya and Karusala, Naveena and DiSalvo, Betsy and Kumar, Neha},
  booktitle={Proceedings of the 2018 Designing Interactive Systems Conference},
  pages={45--58},
  year={2018},
  publisher={ACM}, 
  address={Hong Kong, China}

}

@inproceedings{oguamanam2023intersectional,
  title={An Intersectional Look at Use of and Satisfaction with Digital Mental Health Platforms: A Survey of Perinatal Black Women},
  author={Oguamanam, Vanessa O and Hernandez, Natalie and Chandler, Rasheeta and Guillaume, Dominique and Mckeever, Kai and Allen, Morgan and Mohammed, Sabreen and Parker, Andrea G},
  booktitle={Proceedings of the 2023 CHI Conference on Human Factors in Computing Systems},
  pages={1--20},
  year={2023},
  publisher={ACM},
  address={Hamburg, Germany}
}

@inproceedings{moitra2021negotiating,
  title={Negotiating intersectional non-Normative queer identities in India},
  author={Moitra, Aparna and Marathe, Megh and Ahmed, Syed Ishtiaque and Chandra, Priyank},
  booktitle={Extended Abstracts of the 2021 CHI Conference on Human Factors in Computing Systems},
  pages={1--6},
  year={2021},
  publisher={ACM},
  address={Yokohama, Japan}
}

@article{andalibi2022lgbtq,
  title={LGBTQ Persons’ use of online spaces to navigate conception, pregnancy, and pregnancy loss: an intersectional approach},
  author={Andalibi, Nazanin and Lacombe-Duncan, Ashley and Roosevelt, Lee and Wojciechowski, Kylie and Giniel, Cameron},
  journal={ACM Transactions on Computer-Human Interaction (TOCHI)},
  volume={29},
  number={1},
  pages={1--46},
  year={2022},
  publisher={ACM New York, NY}
}

@article{rankin2019straighten,
  title={Straighten up and fly right: Rethinking intersectionality in HCI research},
  author={Rankin, Yolanda A and Thomas, Jakita O},
  journal={Interactions},
  volume={26},
  number={6},
  pages={64--68},
  year={2019},
  publisher={ACM New York, NY, USA}
}

@article{erete2018intersectional,
  title={An intersectional approach to designing in the margins},
  author={Erete, Sheena and Israni, Aarti and Dillahunt, Tawanna},
  journal={Interactions},
  volume={25},
  number={3},
  pages={66--69},
  year={2018},
  publisher={ACM New York, NY, USA}
}

@inproceedings{sanchez2021framework,
  title={A framework for intersectional perspectives in software engineering},
  author={S{\'a}nchez-Gord{\'o}n, Mary and Colomo-Palacios, Ricardo},
  booktitle={2021 IEEE/ACM 13th International Workshop on Cooperative and Human Aspects of Software Engineering (CHASE)},
  pages={121--122},
  year={2021},
  publisher={IEEE}, 
  address={Madrid, Spain}
}

@inproceedings{mendez2019gendermag,
  title={From GenderMag to InclusiveMag: An inclusive design meta-method},
  author={Mendez, Christopher and Letaw, Lara and Burnett, Margaret and Stumpf, Simone and Sarma, Anita and Hilderbrand, Claudia},
  booktitle={2019 IEEE Symposium on Visual Languages and Human-Centric Computing (VL/HCC)},
  pages={97--106},
  year={2019},
  publisher={IEEE}, 
  address={Memphis, Tennessee, USA}
}

@inproceedings{marsden2016stereotypes,
  title={Stereotypes and politics: reflections on personas},
  author={Marsden, Nicola and Haag, Maren},
  booktitle={Proceedings of the 2016 CHI conference on human factors in computing systems},
  pages={4017--4031},
  year={2016},
  publisher={ACM},
  address={San Jose, CA, USA}
}

@inproceedings{foulds2020intersectional,
  title={An intersectional definition of fairness},
  author={Foulds, James R and Islam, Rashidul and Keya, Kamrun Naher and Pan, Shimei},
  booktitle={2020 IEEE 36th International Conference on Data Engineering (ICDE)},
  pages={1918--1921},
  year={2020},
  publisher={IEEE},
  address={Dallas, TX, USA}
}

@inproceedings{mcdonald2020privacy,
  title={Privacy and power: Acknowledging the importance of privacy research and design for vulnerable populations},
  author={McDonald, Nora and Badillo-Urquiola, Karla and Ames, Morgan G and Dell, Nicola and Keneski, Elizabeth and Sleeper, Manya and Wisniewski, Pamela J},
  booktitle={Extended Abstracts of the 2020 CHI Conference on Human Factors in Computing Systems},
  pages={1--8},
  year={2020}, 
  publisher={ACM},
  address={Honolulu, Hawaii, USA}
}

@inproceedings{klumbyte2022critical,
  title={Critical tools for machine learning: Working with intersectional critical concepts in machine learning systems design},
  author={Klumbytė, Goda and Draude, Claude and Taylor, Alex S},
  booktitle={Proceedings of the 2022 ACM Conference on Fairness, Accountability, and Transparency},
  pages={1528--1541},
  year={2022},
  publisher={ACM},
  address={Seoul, South Korea}
}

@inproceedings{obie2021first,
  title={A first look at human values-violation in app reviews},
  author={Obie, Humphrey O and Hussain, Waqar and Xia, Xin and Grundy, John and Li, Li and Turhan, Burak and Whittle, Jon and Shahin, Mojtaba},
  booktitle={2021 IEEE/ACM 43rd International Conference on Software Engineering: Software Engineering in Society (ICSE-SEIS)},
  pages={29--38},
  year={2021},
  publisher={IEEE}, 
  address={Virtual Event Spain}
}

@inproceedings{tushev2020digital,
  title={Digital discrimination in sharing economy a requirements engineering perspective},
  author={Tushev, Miroslav and Ebrahimi, Fahimeh and Mahmoud, Anas},
  booktitle={2020 IEEE 28th International Requirements Engineering Conference (RE)},
  pages={204--214},
  year={2020},
  publisher={IEEE},
  address={Zurich, Switzerlands}
}

@inproceedings{iqbal2021mining,
  title={Mining reddit as a new source for software requirements},
  author={Iqbal, Tahira and Khan, Moniba and Taveter, Kuldar and Seyff, Norbert},
  booktitle={2021 IEEE 29th international requirements engineering conference (RE)},
  pages={128--138},
  year={2021},
   publisher={IEEE},
  address={Notre Dame, IN, USA}
}

@inproceedings{li2022narratives,
  title={Narratives: the unforeseen influencer of privacy concerns},
  author={Li, Ze Shi and Sihag, Manish and Arony, Nowshin Nawar and Junior, Joao Bezerra and Phan, Thanh and Ernst, Neil and Damian, Daniela},
  booktitle={2022 IEEE 30th International Requirements Engineering Conference (RE)},
  pages={127--139},
  year={2022},
  publisher={IEEE},
 address={Melbourne, Australia}
}

@article{besmer2020investigating,
  title={Investigating user perceptions of mobile app privacy: An analysis of user-submitted app reviews},
  author={Besmer, Andrew R and Watson, Jason and Banks, M Shane},
  journal={International Journal of Information Security and Privacy (IJISP)},
  volume={14},
  number={4},
  pages={74--91},
  year={2020},
  publisher={IGI Global}
}

@inproceedings{shams2020society,
  title={Society-oriented applications development: Investigating users' values from bangladeshi agriculture mobile applications},
  author={Shams, Rifat Ara and Hussain, Waqar and Oliver, Gillian and Nurwidyantoro, Arif and Perera, Harsha and Whittle, Jon},
  booktitle={Proceedings of the ACM/IEEE 42nd International Conference on Software Engineering: Software Engineering in Society},
  pages={53--62},
  year={2020},
publisher={IEEE},
 address={Seoul, South Korea}
}

@article{khalid2014mobile,
  title={What do mobile app users complain about?},
  author={Khalid, Hammad and Shihab, Emad and Nagappan, Meiyappan and Hassan, Ahmed E},
  journal={IEEE software},
  volume={32},
  number={3},
  pages={70--77},
  year={2014},
  publisher={IEEE}
}

@inproceedings{guzman2015ensemble,
  title={Ensemble methods for app review classification: An approach for software evolution (n)},
  author={Guzman, Emitza and El-Haliby, Muhammad and Bruegge, Bernd},
  booktitle={2015 30th IEEE/ACM International Conference on Automated Software Engineering (ASE)},
  pages={771--776},
  year={2015},
  publisher={IEEE},
 address={Lincoln, NE, USA}
}

@inproceedings{guzman2016needle,
  title={A needle in a haystack: What do twitter users say about software?},
  author={Guzman, Emitza and Alkadhi, Rana and Seyff, Norbert},
  booktitle={2016 IEEE 24th international requirements engineering conference (RE)},
  pages={96--105},
  year={2016},
  publisher={IEEE},
 address={Beijing, China}
}

@inproceedings{tabbassum2023towards,
  title={Towards a Cross-Country Analysis of Software-Related Tweets},
  author={Tabbassum, Saliha and Fischer, Ricarda Anna-Lena and Guzman, Emitza},
  booktitle={International Working Conference on Requirements Engineering: Foundation for Software Quality},
  pages={272--282},
  year={2023},
publisher={Springer},
 address={Barcelona, Spain}
}

@article{nayebi2018app,
  title={App store mining is not enough for app improvement},
  author={Nayebi, Maleknaz and Cho, Henry and Ruhe, Guenther},
  journal={Empirical Software Engineering},
  volume={23},
  number={5},
  pages={2764--2794},
  year={2018},
  publisher={Springer}
}

@inproceedings{williams2017mining,
  title={Mining twitter feeds for software user requirements},
  author={Williams, Grant and Mahmoud, Anas},
  booktitle={Proc. of the International Requirements Engineering Conference (RE)},
  pages={1--10},
  year={2017},
publisher={IEEE},
 address={Lisbon, Portugal}
}

@inProceedings{Dennis2013,
author = {Pagano, Dennis and Maalej, Walid},
booktitle = {Proc. of the International Requirements Engineering Conference},
pages = {125--134},
title = {{User feedback in the appstore: an empirical study}},
year = {2013},
publisher={IEEE},
 address={Rio de Janeiro-RJ, Brazil}
}

@inproceedings{guzman2018user,
  title={User feedback in the app store: a cross-cultural study},
  author={Guzman, Emitza and Oliveira, Lu{\'\i}s and Steiner, Yves and Wagner, Laura C and Glinz, Martin},
  booktitle={Proceedings of the 40th International Conference on Software Engineering: Software Engineering in Society},
  pages={13--22},
  year={2018},
publisher={IEEE},
 address={Gothenburg, Sweden}
}

@inproceedings{fischer2021does,
  title={Does culture matter? impact of individualism and uncertainty avoidance on app reviews},
  author={Fischer, Ricarda Anna-Lena and Walczuch, Rita and Guzman, Emitza},
  booktitle={2021 IEEE/ACM 43rd International Conference on Software Engineering: Software Engineering in Society (ICSE-SEIS)},
  pages={67--76},
  year={2021},
  publisher={IEEE},
 address={Virtual Event Spain}
}

@inproceedings{guzman2017little,
  title={A little bird told me: Mining tweets for requirements and software evolution},
  author={Guzman, Emitza and Ibrahim, Mohamed and Glinz, Martin},
  booktitle={2017 IEEE 25th International requirements engineering conference (RE)},
  pages={11--20},
  year={2017},
  publisher={IEEE},
 address={Lisbon, Portugal}
}

@inproceedings{maalej2015bug,
  title={Bug report, feature request, or simply praise? on automatically classifying app reviews},
  author={Maalej, Walid and Nabil, Hadeer},
  booktitle={2015 IEEE 23rd international requirements engineering conference (RE)},
  pages={116--125},
  year={2015},
  publisher={IEEE},
 address={Ottawa, ON, Canada}
}

@inproceedings{panichella2015can,
  title={How can i improve my app? classifying user reviews for software maintenance and evolution},
  author={Panichella, Sebastiano and Di Sorbo, Andrea and Guzman, Emitza and Visaggio, Corrado A and Canfora, Gerardo and Gall, Harald C},
  booktitle={2015 IEEE international conference on software maintenance and evolution (ICSME)},
  pages={281--290},
  year={2015},
  publisher={IEEE},
 address={Bremen, Germany}
}

@article{hoon2013analysis,
author = {Hoon, Leonard and Vasa, Rajesh and Schneider, Jean-Guy and Grundy, John and Others},
journal = {Swinburne University of Technology, Tech. Rep},
title = {{An analysis of the mobile app review landscape: trends and implications}},
year = {2013}
}

@inproceedings{gao2018online,
  title={Online app review analysis for identifying emerging issues},
  author={Gao, Cuiyun and Zeng, Jichuan and Lyu, Michael R and King, Irwin},
  booktitle={Proceedings of the 40th International Conference on Software Engineering},
  pages={48--58},
  year={2018},
publisher={IEEE},
 address={Gothenburg, Sweden}
}

@inproceedings{stronstads,
  title={What’s next in my backlog? Time series analysis of user reviews},
  author={Str{\o}nstad, G{\o}ran H and Gerostathopoulos, Ilias and Guzm{\'a}n, Emitz{\'a}},
  booktitle={Proceedings of the Workshop on Empirical Requirements Engineering},
  year={2023}
}

@misc{Benner_Thrush_Isaac_2019, 
title={Facebook engages in housing discrimination with its AD practices, U.S. says}, url={https://www.nytimes.com/2019/03/28/us/politics/facebook-housing-discrimination.html}, 
journal={The New York Times}, 
publisher={The New York Times}, 
author={Benner, Katie and Thrush, Glenn and Isaac, Mike}, 
year={2019}, 
month={Mar}}

@misc{year_in_review, 
title={The year in review: Top news stories of 2022 month-by-month},
url={https://www.cbsnews.com/news/the-year-in-review-top-news-stories-of-2022-month-by-month/},
author={Pauley, Jane and Morgan, David}, 
journal={CBS News}, 
publisher={CBS News},
year={2023},
month={Jan}
}

@misc{Biron_2022, title={Facebook sued over death of federal officer}, url={https://www.nbcnews.com/tech/internet/facebook-sued-death-federal-officer-rcna11098}, journal={NBCNews.com}, publisher={NBCUniversal News Group}, author={Biron, Bethany}, year={2022}, month={Jan}}

@misc{Clayton_2021, title={Rohingya sue Facebook for \$150bn over Myanmar hate speech}, url={https://www.bbc.com/news/world-asia-59558090}, journal={BBC News}, publisher={BBC}, author={Clayton, James}, year={2021}, month={Dec}}

@misc{Thorbecke_2020, title={Facebook hit with lawsuit over Kenosha protest deaths}, url={https://abcnews.go.com/US/facebook-hit-lawsuit-kenosha-protest-deaths/story?id=73189351}, journal={ABC News}, publisher={ABC News Network}, author={Thorbecke, Catherine}, year={2020}, month={Sep}}

@misc{Whitcomb_2017, title={Families of San Bernardino shooting sue facebook, Google, Twitter}, url={https://www.reuters.com/article/us-sanbernardino-attack-lawsuit-idUSKBN1802SL}, journal={Reuters}, publisher={Thomson Reuters}, author={Whitcomb, Dan}, year={2017}, month={May}}

@article{Taylor18prophet,
author = {Sean J. Taylor and Benjamin Letham},
title = {Forecasting at Scale},
journal = {The American Statistician},
volume = {72},
number = {1},
pages = {37-45},
year = {2018},
publisher = {Taylor & Francis},
doi = {10.1080/00031305.2017.1380080},
URL = {https://doi.org/10.1080/00031305.2017.1380080},
eprint = {https://doi.org/10.1080/00031305.2017.1380080}
}

@misc{hyndman23forecast,
title = {{forecast}: Forecasting functions for time series and linear models},
author = {Rob Hyndman and George Athanasopoulos and Christoph  Bergmeir and Gabriel Caceres and Leanne Chhay and Mitchell O'Hara-Wild and Fotios Petropoulos and Slava Razbash and Earo Wang and Farah Yasmeen},
year = {2023},
note = {R package version 8.21.1},
url = {https://pkg.robjhyndman.com/forecast/},

}

@Article{hyndman08forecast,
title = {Automatic time series forecasting: the forecast package for {R}},
author = {Rob J Hyndman and Yeasmin Khandakar},
journal = {Journal of Statistical Software},
volume = {26},
number = {3},
pages = {1--22},
year = {2008},
doi = {10.18637/jss.v027.i03},
}

@ARTICLE{9952173,
  author={Gao, Cuiyun and Li, Yaoxian and Qi, Shuhan and Liu, Yang and Wang, Xuan and Zheng, Zibin and Liao, Qing},
  journal={IEEE Transactions on Reliability}, 
  title={Listening to Users' Voice: Automatic Summarization of Helpful App Reviews}, 
  year={2023},
  volume={72},
  number={4},
  pages={1619-1631},
  doi={10.1109/TR.2022.3217566}}

@inproceedings{chen2014ar,
  title={AR-miner: mining informative reviews for developers from mobile app marketplace},
  author={Chen, Ning and Lin, Jialiu and Hoi, Steven CH and Xiao, Xiaokui and Zhang, Boshen},
  booktitle={Proceedings of the 36th international conference on software engineering},
  pages={767--778},
  year={2014}
}

@article{mccabe2025estimating,
  title={Estimating substance use disparities across intersectional social positions using machine learning: An application of group-lasso interaction network.},
  author={McCabe, Connor J and Helm, Jonathan L and Halvorson, Max A and Blaikie, Kieran J and Lee, Christine M and Rhew, Isaac C},
  journal={Psychology of Addictive Behaviors},
  volume={39},
  number={2},
  pages={113},
  year={2025},
  publisher={American Psychological Association}
}

@article{tibshirani1996regression,
  title={Regression shrinkage and selection via the lasso},
  author={Tibshirani, Robert},
  journal={Journal of the Royal Statistical Society Series B: Statistical Methodology},
  volume={58},
  number={1},
  pages={267--288},
  year={1996},
  publisher={Oxford University Press}
}

@article{olivier2017relative,
  title={Relative effect sizes for measures of risk},
  author={Olivier, Jake and May, Warren L and Bell, Melanie L},
  journal={Communications in statistics-theory and methods},
  volume={46},
  number={14},
  pages={6774--6781},
  year={2017},
  publisher={Taylor \& Francis}
}

@article{hecht2021sample,
  title={s},
  author={Hecht, Martin and Zitzmann, Steffen},
  journal={Structural Equation Modeling: A Multidisciplinary Journal},
  volume={28},
  number={2},
  pages={229--236},
  year={2021},
  publisher={Taylor \& Francis}
}

@misc{analyticsfacebook2025,
	title        = {How Facebook leverages Large Language Models to understand user bug reports and guide fundamental improvements},
	author       = {Analytics at Meta},
	year         = {2025},
	publisher    = {Medium},
	howpublished = {\url{https://medium.com/\@AnalyticsAtMeta/how-facebook-leverages-large-language-models-to-understand-user-bug-reports-and-guide-fundamental-70ab26475850}}
}

@misc{sprinklrmicrosoft2025,
	title        = {Microsoft's social intelligence team turns VoC data into actionable insights},
	author       = {Sprinklr},
	year         = {2025},
	publisher    = {Sprinklr},
	howpublished = {\url{https://www.sprinklr.com/stories/microsoft}}
}

@article{truong2020selective,
  title={Selective review of offline change point detection methods},
  author={Truong, Charles and Oudre, Laurent and Vayatis, Nicolas},
  journal={Signal processing},
  volume={167},
  pages={107299},
  year={2020},
  publisher={Elsevier}
}

@article{gasparrini2010distributed,
  title={Distributed lag non-linear models},
  author={Gasparrini, Antonio and Armstrong, Ben and Kenward, Mike G},
  journal={Statistics in medicine},
  volume={29},
  number={21},
  pages={2224--2234},
  year={2010},
  publisher={Wiley Online Library}
}

@misc{applellm2024,
	title        = {An LLM-Based Approach to Review Summarization on the App Store},
	author       = {Apple Machine Learning Research},
	year         = {2025},
	publisher    = {Apple},
	howpublished = {\url{https://machinelearning.apple.com/research/app-store-review}}
}

@misc{kearnsnew2025,
	title        = {New Play Store summaries save you from skimming app reviews},
	author       = {Taylor Kearns and AssembleDebug},
	year         = {2025},
	publisher    = {Android Authority},
	howpublished = {\url{https://www.androidauthority.com/play-store-ai-generated-review-summaries-3611995/}}
}

@article{gokgoz2025if,
  title={If it ain’t broke, should you still fix it? effects of incorporating user feedback in product development on mobile application ratings},
  author={Gokgoz, Zeynep Aydin and Ataman, M Berk and Van Bruggen, Gerrit H},
  journal={International Journal of Research in Marketing},
  volume={42},
  number={2},
  pages={467--486},
  year={2025},
  publisher={Elsevier}
}

@book{hosmer2013applied,
  title={Applied logistic regression},
  author={Hosmer Jr, David W and Lemeshow, Stanley and Sturdivant, Rodney X},
  year={2013},
  publisher={John Wiley \& Sons}
}

@article{lim2015learning,
  title={Learning interactions via hierarchical group-lasso regularization},
  author={Lim, Michael and Hastie, Trevor},
  journal={Journal of Computational and Graphical Statistics},
  volume={24},
  number={3},
  pages={627--654},
  year={2015},
  publisher={Taylor \& Francis}
}

@article{nielsen2025intersectional,
  title={Intersectional analysis for science and technology},
  author={Nielsen, Mathias Wullum and Gissi, Elena and Heidari, Shirin and Horton, Richard and Nadeau, Kari C and Ngila, Dorothy and Noble, Safiya Umoja and Paik, Hee Young and Tadesse, Girmaw Abebe and Zeng, Eddy Y and others},
  journal={Nature},
  volume={640},
  number={8058},
  pages={329--337},
  year={2025},
  publisher={Nature Publishing Group UK London}
}

@article{dkabrowski2022analysing,
  title={Analysing app reviews for software engineering: a systematic literature review},
  author={D{\k{a}}abrowski, Jacek and Letier, Emmanuel and Perini, Anna and Susi, Angelo},
  journal={Empirical Software Engineering},
  volume={27},
  number={2},
  pages={43},
  year={2022},
  publisher={Springer}
}

@inproceedings{licorish2017attributes,
  title={Attributes that predict which features to fix: Lessons for app store mining},
  author={Licorish, Sherlock A and Savarimuthu, Bastin Tony Roy and Keertipati, Swetha},
  booktitle={Proceedings of the 21st International Conference on Evaluation and Assessment in Software Engineering},
  pages={108--117},
  year={2017}
}

@article{malgaonkar2022prioritizing,
  title={Prioritizing user concerns in app reviews--A study of requests for new features, enhancements and bug fixes},
  author={Malgaonkar, Saurabh and Licorish, Sherlock A and Savarimuthu, Bastin Tony Roy},
  journal={Information and Software Technology},
  volume={144},
  pages={106798},
  year={2022},
  publisher={Elsevier}
}

@inproceedings{shahin2023study,
  title={A study of gender discussions in mobile apps},
  author={Shahin, Mojtaba and Zahedi, Mansooreh and Khalajzadeh, Hourieh and Nasab, Ali Rezaei},
  booktitle={2023 IEEE/ACM 20th International Conference on Mining Software Repositories (MSR)},
  pages={598--610},
  year={2023},
  organization={IEEE}
}

@article{kirk2021bias,
  title={Bias out-of-the-box: An empirical analysis of intersectional occupational biases in popular generative language models},
  author={Kirk, Hannah Rose and Jun, Yennie and Volpin, Filippo and Iqbal, Haider and Benussi, Elias and Dreyer, Frederic and Shtedritski, Aleksandar and Asano, Yuki},
  journal={Advances in neural information processing systems},
  volume={34},
  pages={2611--2624},
  year={2021}
}

@inproceedings{buolamwini2018gender,
  title={Gender shades: Intersectional accuracy disparities in commercial gender classification},
  author={Buolamwini, Joy and Gebru, Timnit},
  booktitle={Conference on fairness, accountability and transparency},
  pages={77--91},
  year={2018},
  publisher={PMLR},
  address={New York, NY, USA}
}

@inproceedings{cabrera2019fairvis,
  title={FairVis: Visual analytics for discovering intersectional bias in machine learning},
  author={Cabrera, {\'A}ngel Alexander and Epperson, Will and Hohman, Fred and Kahng, Minsuk and Morgenstern, Jamie and Chau, Duen Horng},
  booktitle={2019 IEEE Conference on Visual Analytics Science and Technology (VAST)},
  pages={46--56},
  year={2019},
  publisher={IEEE},
  address={Vancouver, BC, Canada}
}

@inproceedings{guo2021detecting,
  title={Detecting emergent intersectional biases: Contextualized word embeddings contain a distribution of human-like biases},
  author={Guo, Wei and Caliskan, Aylin},
  booktitle={Proceedings of the 2021 AAAI/ACM Conference on AI, Ethics, and Society},
  pages={122--133},
  year={2021},
  publisher={ACM},
  address={Virtual Event USA}
}

@inproceedings{ghai2021wordbias,
  title={Wordbias: An interactive visual tool for discovering intersectional biases encoded in word embeddings},
  author={Ghai, Bhavya and Hoque, Md Naimul and Mueller, Klaus},
  booktitle={Extended Abstracts of the 2021 CHI Conference on Human Factors in Computing Systems},
  pages={1--7},
  year={2021},
  publisher={ACM},
  address={Yokohama, Japan}
}

@article{tan2019assessing,
  title={Assessing social and intersectional biases in contextualized word representations},
  author={Tan, Yi Chern and Celis, L Elisa},
  journal={Advances in neural information processing systems},
  volume={32},
  year={2019}
}

@inproceedings{steed2021image,
  title={Image representations learned with unsupervised pre-training contain human-like biases},
  author={Steed, Ryan and Caliskan, Aylin},
  booktitle={Proceedings of the 2021 ACM conference on fairness, accountability, and transparency},
  pages={701--713},
  year={2021},
  publisher={ACM},
  address={Virtual Event Canada}
}

\end{document}